\documentclass[
 reprint,
superscriptaddress,
amsmath,amssymb,showkeys
]{revtex4-2}
\usepackage{float} 
\usepackage{graphicx}
\usepackage{dcolumn}
\usepackage{bm}
\usepackage{multirow}
\usepackage{longtable}
\usepackage[a4paper, left=.5in, right=.7in, textwidth=7.8in]{geometry}

\usepackage{booktabs}

\usepackage{array}

\usepackage{hyperref}  
\usepackage{cleveref}
\makeatletter
\def\blfootnote{\gdef\@thefnmark{}\@footnotetext}
\makeatother

\usepackage{titlesec}
\titleformat{\paragraph}[runin]{\bfseries\itshape}{\textbf{\textit{\theparagraph.}}}{1em}{}
\titleformat{\subparagraph}[runin]{\bfseries\itshape}{\textbf{\textit{\thesubparagraph.}}}{1em}{}[:]

\renewcommand{\theparagraph}{\thesection\alph{paragraph}}  
\renewcommand{\thesubparagraph}{\theparagraph.\arabic{subparagraph}} 

\begin{document}

\preprint{}

\title{\Large Oscillatory dynamics between language usage and economic activity}


\author{Alejandro Pardo Pintos}
\affiliation{Universidad de Buenos Aires, Facultad de Ciencias Exactas y Naturales, Departamento de Física. Buenos Aires, Argentina.}
\affiliation{CONICET - Universidad de Buenos Aires, Instituto de Física Interdisciplinaria y Aplicada (INFINA). Buenos Aires, Argentina.}
\author{Diego E Shalom}
\affiliation{Universidad de Buenos Aires, Facultad de Ciencias Exactas y Naturales, Departamento de Física. Buenos Aires, Argentina.}
\affiliation{CONICET - Universidad de Buenos Aires, Instituto de Física Interdisciplinaria y Aplicada (INFINA). Buenos Aires, Argentina.}
\author{Guillermo Cecchi}
\affiliation{IBM Research, Yorktown Heights, NY, USA.}
\author{Gabriel Mindlin}
\affiliation{Universidad de Buenos Aires, Facultad de Ciencias Exactas y Naturales, Departamento de Física. Buenos Aires, Argentina.}
\affiliation{CONICET - Universidad de Buenos Aires, Instituto de Física Interdisciplinaria y Aplicada (INFINA). Buenos Aires, Argentina.}
\author{Marcos A Trevisan}
\email{marcos@df.uba.ar}
\affiliation{Universidad de Buenos Aires, Facultad de Ciencias Exactas y Naturales, Departamento de Física. Buenos Aires, Argentina.}
\affiliation{CONICET - Universidad de Buenos Aires, Instituto de Física Interdisciplinaria y Aplicada (INFINA). Buenos Aires, Argentina.}

\date{\today}


\begin{abstract}
\noindent Over the past two centuries, the frequency of word usage in major Western languages has exhibited small amplitude regular cycles, superimposed on larger background trends. We show that these cycles of word usage organize into semantically coherent clusters that track, interact with, or precede macroeconomic rhythms. To explore the nature of this interaction, we build on a minimal mathematical model that captures usage dynamics through parameters linked to real world processes. Parameter fitting reveals that word usage operates near a Hopf bifurcation, a critical regime associated with heightened sensitivity, uncovering a robust and specific coupling between language usage and economic activity. In linguistic datasets that allow long term quantitative analysis, rooted in capitalist contexts, this coupling appears as an oscillatory dynamic, reflecting a feedback process in which language both shapes material reality and manages the changes it produces.
\end{abstract}


\maketitle

\blfootnote {Author contributions: 
GC, GM and MAT designed research;
APP, DES, and MAT performed research; 
APP and DES analyzed data; 
and APP, DES, GC, GM and MAT wrote the paper. The authors declare no conflict of interest.}

\noindent {\bf Significance.}
Analyses of large linguistic datasets often focus either on structural patterns in language or on sociocultural content. Here we integrate both perspectives, showing that word usage exhibits cycles that synchronize into semantic communities, which precede or track cycles of macroeconomic activity. We show that language cycles operate near a critical point that enhances sensitivity to changes. These findings support the view that language shapes the material reality and manage the changes it produces, generating cycles from this feedback process. In datasets that allow long term quantitative analysis, which are rooted in historically capitalist contexts, this feedback emerges as an oscillatory dynamic between language usage and economic activity.

\section*{\large Introduction}

Zipf's law is a well-known empirical regularity of language, stating that the frequency of a word is inversely proportional to its rank, generally following a power-law distribution \cite{Zipf1933}. However, word usage changes over time. For example, vocabulary related to religion was more common two centuries ago. As the trends of these words declined in usage, others became more prominent, but Zipf’s law still applies \cite{aitchison2016zipf}.
Recent analyses of large corpora have revealed small-amplitude oscillations superimposed on the trends in noun frequency \cite{Montemurro2016,pintos2022cognitive}. In this study, we explore these oscillations, consistently observed across multiple languages over the past two centuries.

The selection of languages and the analysis period that yielded most of these results were guided by the largest available linguistic corpora \cite{Michel2010, davies2010corpus}, which provide statistical robustness, enhance linguistic diversity, and help mitigate dataset biases \cite{Pechenick2015,Younes2019}. However, these choices also reflect genealogical and cultural connections among the languages studied, particularly their shared historical exposure to capitalism \cite{lansdall2017content}.
Interestingly, the regular oscillations in noun usage across Western languages closely match the rhythm of the paradigmatic Juglar or business cycles, characteristic of modern industrial and postindustrial economies \cite{juglar1862}.  

Business cycles, typically lasting 7 to 12 years \cite{kurkina2017mathematical}, have been described in neoclassical terms  as follows: in the early years of the upswing phase, the excess demand over supply cannot be met by the existing fixed capital. As a result, new capital assets must be created through increased investments. When demand begins to decline, the impact on output is delayed, even if the initial growth in output was driven by the increased utilization of existing capital \cite{Schumpeter1939}. This description enables the modeling of business cycles using dynamical systems with delay \cite{kaleki1935macrodynamic, kurkina2017mathematical, cai2005hopf}.
Other relevant economic cycles can be explained on the same basis: Kitchin's cycles arise from inventory and information lags in production and distribution processes, making them faster, with periods of around 3 to 5 years \cite{kitchin1923}; on the other side, Kuznets' swings reflect long-term structural changes such as demographic shifts and infrastructure investment, extending the cycle period to 15–25 years \cite{kuznets1958quantitative}. These families of cycles have been confirmed by spectral analysis of global Gross Domestic Product (GDP), a key indicator of macroeconomic activity over the past 150 years \cite{korotayev2010spectral}.

\begin{figure*}[!htbp]
\centering \includegraphics[width=12cm]{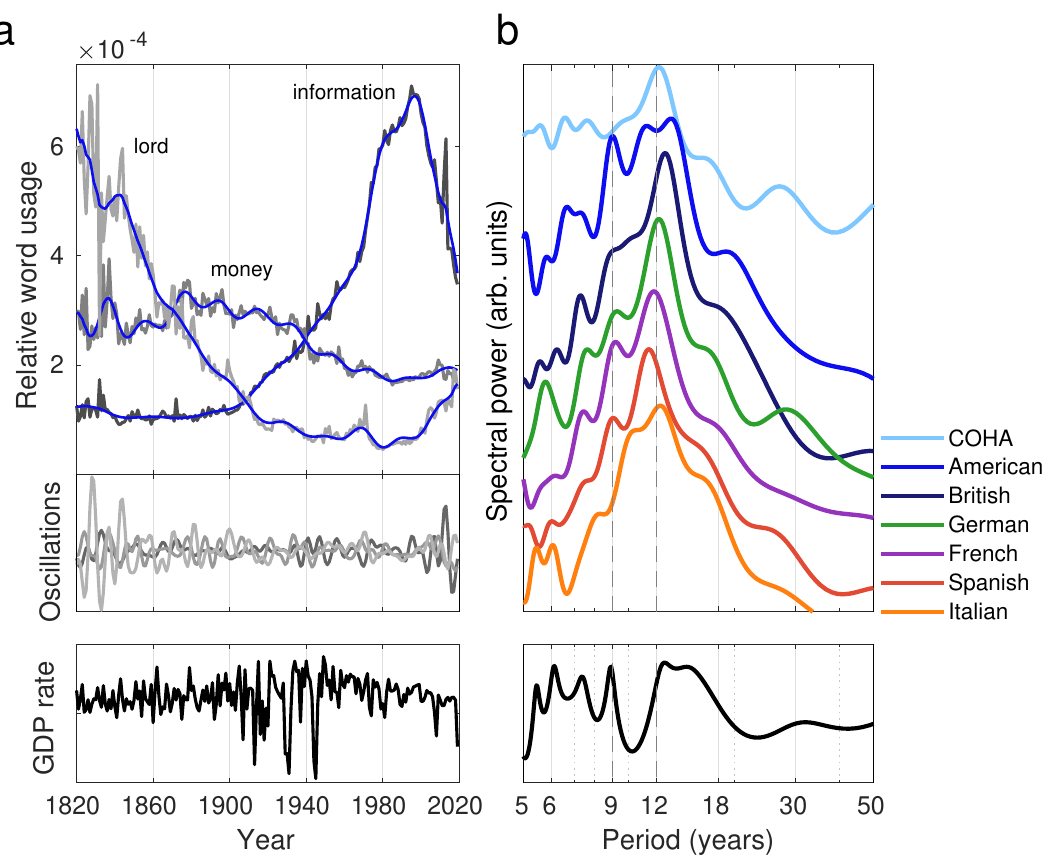}
\caption{\textbf{Word usage exhibits small-amplitude oscillations aligned with business cycles.} \textbf{(a)} Time series of relative usage for the words {\em lord, money}, and {\em information} using Google Books American. After subtracting the large trend components, small-amplitude perturbations in word usage become apparent, as shown in the middle panel. Time series of the rate of change in global Gross Domestic Product is also shown in black in the bottom panel \textbf{(b)} Spectral analyses of the small-amplitude components reveal a rich spectrum, with dominant periods of 9 and 12 years across languages and corpora, and smaller, language-dependent peaks. In the bottom panel we show the spectral components of the rate of change in GDP, which also shows peaks at 9 and 12 years.}
\label{fig:time_series}
\end{figure*}

The alignment between oscillations in word usage and GDP fluctuations invites the question of whether language merely reflects material conditions or may also anticipate or participate in shaping them.
Here, we show synchronized word oscillations form semantic communities, some of which are linked to macroeconomic activity, either tracking or anticipating shifts in global GDP. We further explore the nature of this relationship using a minimal delay model \cite{pintos2022cognitive} that captures essential features of word dynamics using parameters that both reflect real-world processes and fit empirical word usage time series spanning two centuries and multiple languages. Model fitting suggests that the word usage operates near a Hopf bifurcation, a critical regime indicating a tight relationship between linguistic and economic cycles.

\section*{\large Results} 

We compiled the most stable nouns from the past two centuries across the largest Google Books databases: German (5,987 nouns), French (5,336 nouns), Spanish (3,405 nouns), Italian (3,155 nouns), and, for cross-corpus comparability, the most frequent nouns shared between American and British English, using data from Google Books and the Corpus of Historical American English (COHA), resulting in a total of 4,958 nouns (see Methods).

For each language and year, we calculated the relative usage of a word as the ratio of occurrences to total word count. Examples for the words \textit{lord}, \textit{money}, and \textit{information} from Google American are shown in Figure~\ref{fig:time_series}a. For simplicity, we refer to the relative usage of a word as word usage throughout this work.

\begin{figure*}[ht]
  \centering \includegraphics[width=18cm]
{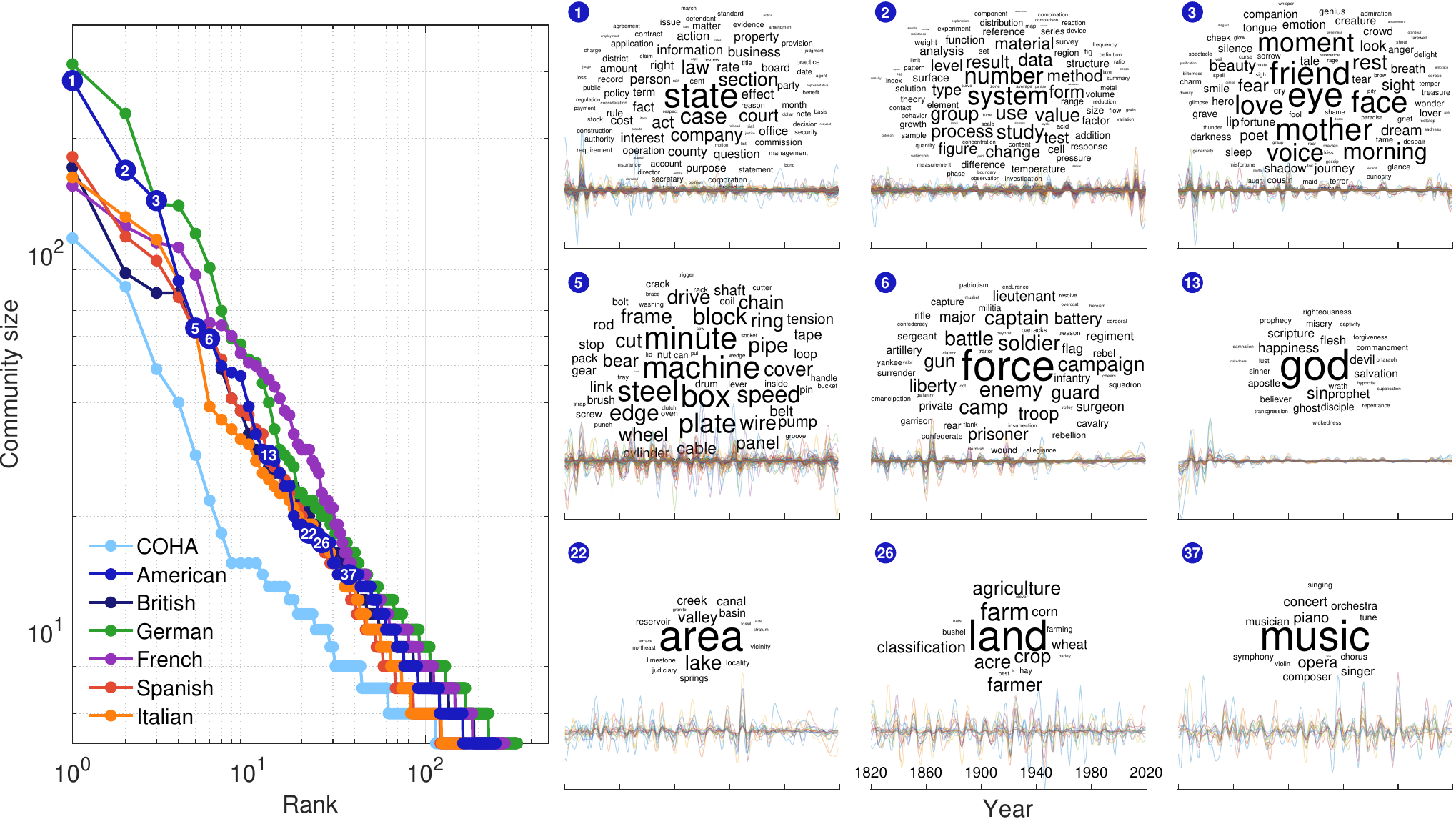}\caption{\textbf{Synchronous words form semantic communities.} Power law exponents are consistent across languages: $\alpha = -0.76 \pm 0.01$, $R^2 = 0.98$ (American English);  $\alpha = -0.67 \pm 0.01$, $R^2 = 0.99$ (British English); $\alpha = -0.78 \pm 0.02$, $R^2 = 0.98$ (German); $\alpha = -0.77 \pm 0.01$, $R^2 = 0.98$ (French); $\alpha = -0.79 \pm 0.01$, $R^2 = 0.99$ (Spanish); and $\alpha = -0.75 \pm 0.01$, $R^2 = 0.99$ (Italian). COHA is an exception, with $\alpha = -0.56 \pm 0.02$, $R^2 = 0.94$.
Larger communities (60–300 words) include domains related to government, science, body parts, family, mechanics, military, and religion. Mid-sized communities reflect more specific fields such as geography, agriculture, and music.}
\label{fig:communities}
\end{figure*}

We isolated long-term trends using Singular Spectrum Analysis (SSA), a data-driven method that makes no assumptions about the underlying structure of the time series. After removing trends, small amplitude fluctuations became apparent (Figure~\ref{fig:time_series}a, mid panel). A Linear Predictive Coding (LPC) spectral analysis of these fluctuations (Figure~\ref{fig:time_series}b) revealed prominent peaks at 12- and 9-year periods, consistently observed across languages, along with secondary language-specific components (see Methods). Compatible with other studies \cite{Montemurro2001,pintos2022cognitive}, these results point to robust, small-amplitude oscillations in word usage with characteristic periodicities between 9 and 12 years, shared across linguistic corpora.

\begin{figure*}[ht]
  \centering
  \includegraphics[width=11cm]{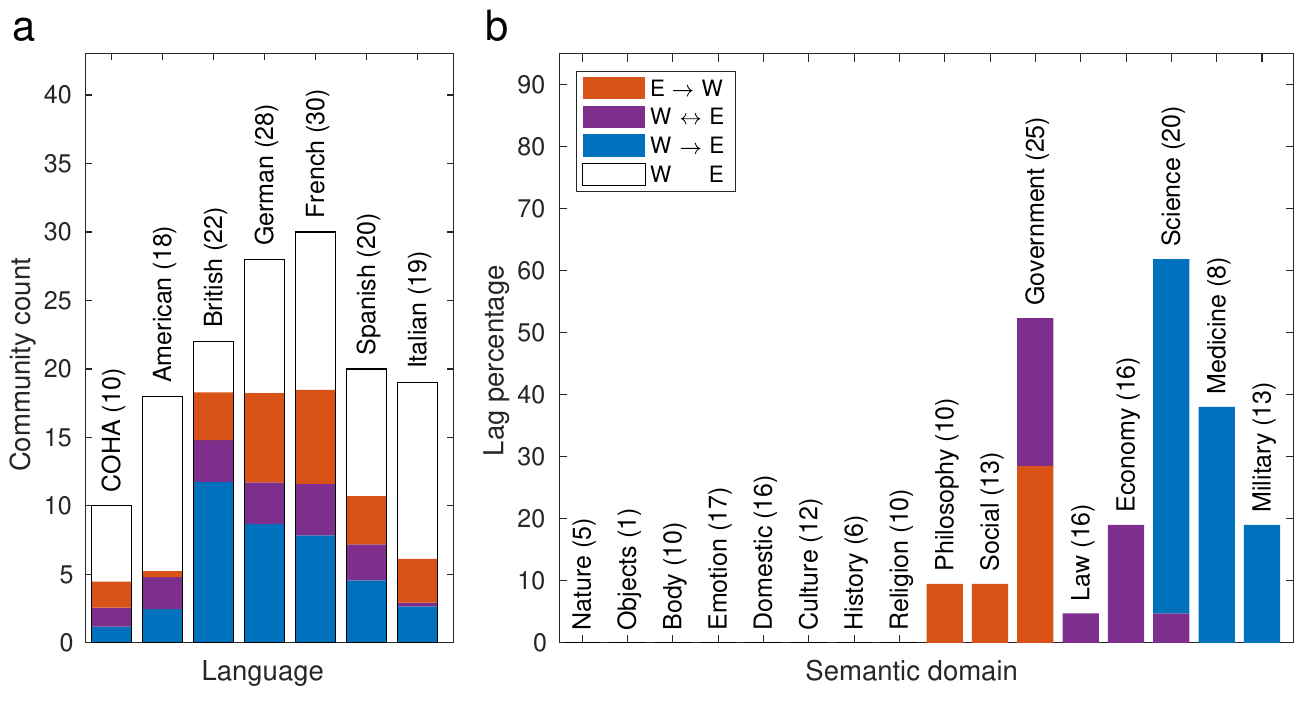}
\caption{\textbf{Semantic fields track or follow macroeconomic activity.} Granger causality tests with time lags used to determine the directionality between language and economic activity across languages. Word communities tracking GPD economic activity ($E\rightarrow W$) are shown in red, those preceding economic activity ($W\rightarrow E$) in blue, and the bidirectional ($W\leftrightarrow E$) in purple; unrelated pairs remained uncolored.
\textbf{(a)} Each database shows a mix of directionalities, with a higher overall proportion of unrelated and GDP-driving (blue) communities.
\textbf{(b)} Communities related to nature, objects, human body, domestic life, emotion, and religion remained unrelated to GDP dynamics, consistent with their lower affinity to economic activity. In contrast, economically relevant linguistic domains were significantly associated with GDP variation. GDP dynamics tends to predict shifts in social and phylosophical terms, while bidirectional feedback appears with economic terms. Government terms both track GDP and present bidirectional links, depending on the lag. Science, medical and military terms consistently precede GDP changes.}
\label{fig:granger}
\end{figure*}

\begin{figure*}[ht]
  \centering
\includegraphics[width=13cm]{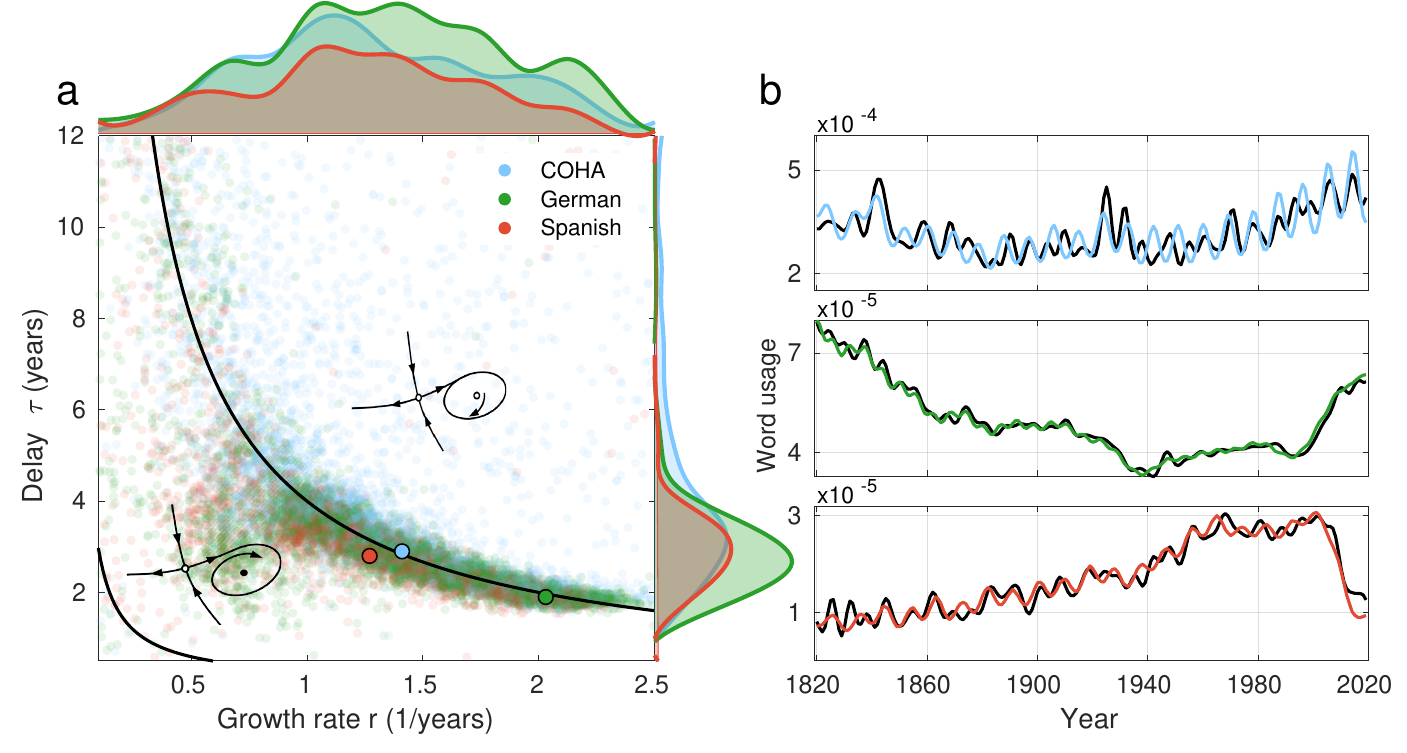}
  \caption{\textbf{Word usage is poised at dynamical criticality}. \textbf{(a)} The logistic model with delay shows a Hopf bifurcation (upper black line) at $\tau=4/r$. Above that curve, oscillations are sustained. Points show the best fits for words in COHA, Google Books German, and Google Books Spanish datasets respectively. The words cluster along the Hopf bifurcation.  Mean fit parameters for each dataset are: COHA, $r = 1.3\pm 0.6~~1/\text{years}, \tau = 4\pm2$ years; German $r = 1.3\pm 0.5~~1/\text{years}, \tau = 3\pm2$ years; Spanish $r = 1.2\pm 0.5~~1/\text{years}, \tau = 3\pm1$ years. \textbf{(b)} Examples of experimental and model-fitted time series for three nouns of the displayed languages: \textit{body}, \textit{aufmerksamkeit} (attention), and \textit{crítico} (critic). Model fitting for each language can be found in Supplementary Figure \ref{fig:sup_hopf}}.
\label{fig:hopf}
\end{figure*}

To investigate external correlates of these linguistic oscillations, we extracted global and country-level GDP time series from 1820 to 2019 and computed their rate of change (see Methods). The bottom panels of Figure~\ref{fig:time_series} display the time series of global GDP rate of change  along with its LPC spectrum. Notably, the  spectrum exhibits energy concentrated in frequency bands corresponding to the largest spectral peaks observed in word usage, which are associated with business cycles, suggesting a potential link between them. Additional GDP peaks are also observed, including those corresponding to Kitchin's cycles (5–6 years) and Kuznets' swings (15–25 years). The spectral components of the latter partially overlap with those of the business cycles, though the spectral load across languages diminishes.

\noindent{\bf Communities of synchronous nouns.} Words do not oscillate in isolation; rather, they form synchronic communities that we identified using the Louvain algorithm with phase coherence as a clustering metric. The algorithm  works by repeatedly grouping nodes into communities that maximize a quantity called modularity, essentially grouping nodes that oscillate strongly in phase with each other, but weakly to the rest of the network. The Louvain algorithm incorporates a resolution parameter that influences community size and number. To calibrate this parameter specifically for language dynamics, we capitalized on a well-established linguistic property known as Heaps' law \cite{petersen2012languages}, which describes how vocabulary grows as language evolves. In this process, stable nouns undergo very few (if any) semantic shifts, while continuing to play an integral role in communication. For example, the dominant meaning of {\em plane} was geometric, before it shifted in the early twentieth century to refer to an aircraft. {\em Monitor} originally referred to someone who issued warnings, before acquiring its modern sense around the mid twentieth century, as a device or system that tracks something for a specific purpose. We fine tuned the resolution parameter until a single value captured documented semantic shifts reported in the Oxford English Dictionary across English corpora (see Methods). In this way, the communities across time reflected these changes, as shown in Figure~\ref{fig:communities_time} (the complete set is available in Table~\ref{table:meanings}).

When the clustering is applied to words that have maintained consistent synchrony over the past two centuries, we obtain communities that reflect persistent core vocabularies. These word communities exhibit a scale-free size distribution with exponents around $0.7$, as shown in Figure~\ref{fig:communities}, along with representative communities from semantic domains observed across languages (see Table~\ref{table:communities} for the complete list). The 10 largest communities span broad semantic domains such as economy, law, government, science, emotions, and human relationships. Other large communities correspond to more specific domains, including medicine, warfare, and mechanics. Smaller communities (ranks 10 to 100) include terms related to religion, agriculture, music, geography, and others.

\noindent\textbf{Directionality between language and economy.} Spectral analysis of word usage and GDP suggests a relationship between economic and linguistic dynamics. Specifically, the largest peaks at 9- and 12-years, common across languages, correspond to the typical periodicity of the well-known business cycles.

We characterized the dynamics of a word community as the average of the individual time series of the words it contains, as illustrated by the examples below the word clouds in Figure~\ref{fig:communities}. We then applied Granger causality to examine the relationship between each community and economic activity, measured as the rate of change of GDP. This method tests whether past values of one time series improve the prediction of another, allowing us to identify leading and lagging dynamics between the two processes. For instance, if the past values of a word community improve the prediction of GDP beyond what its own past explains, we say that the word community precedes economic activity ($W\rightarrow E$). This framework enables us to determine whether a community's dynamics precede, follow, interact with, or are unrelated to economic activity, thus revealing temporal directionality between the two processes (see Methods).

We conducted Granger tests across a range of lags, spanning approximately three full cycles of influence. Figure~\ref{fig:granger} summarizes the results, showing directionality averaged across lags. Word communities that precede GDP economic activity ($W \rightarrow E$) are shown in blue; economic activity preceding word usage ($E \rightarrow W$) in red; feedback cases ($W \leftrightarrow E$) in purple; and unrelated cases in white. As shown in Figure~\ref{fig:granger}a, each database contains a mixture of all four categories, with an average of 45\% of communities showing no significant interaction. Among the remaining communities, which exhibit directional coupling, the majority correspond to $W \rightarrow E$ (27\%), followed by $E \rightarrow W$ (17\%) and bidirectional feedback $W \leftrightarrow E$ (11\%).

We then  identified a set of semantic domains that encompass all the communities across languages (see Methods). Communities related to nature, objects, the human body, domestic life, emotion, and religion are unrelated to economic dynamics (Figure \ref{fig:granger}b). In contrast, domains more closely tied to collective or institutional activity are associated with economic dynamics. Specifically, there is a tendency for economic dynamics to precede the dynamics of social and philosophical terms, and bidirectional feedback is observed with economic and legal terms. The dynamics of government-related terms are generally preceded by economic activity, though reciprocal influence is also present depending on the lag. Notably, a strong association is found for science-related terms, which consistently precede economic variation. Similar anticipatory patterns are observed for terms related to military and medicine.

\noindent {\bf Dynamical model of word usage.} To further study the nature of the relationship between language usage and economy, we build on a minimal model that captures the key features of the observed word dynamics, using a few general parameters that link directly to real-world linguistic and economic processes.

In previous work \cite{pintos2022cognitive}, a logistic delay model was used to describe word usage. In this framework, perturbations in word usage grow at a rate $r$ and are inhibited by sustained past consumption, with a mean delay $\tau$ (see equations in Methods). When these two forces balance, the equilibrium equals the observed long-term trend. This equilibrium remains stable for short delays but becomes unstable beyond a critical delay, $\tau=4/r$ (black curve), where a Hopf bifurcation leads to sustained oscillations (Figure~\ref{fig:hopf}a). Parameters obtained from best fits to empirical time series are tuned along this Hopf bifurcation across languages. Under the strict fitting criteria applied, an average of 83\% of the nouns were found in the vicinity of the bifurcation, with a mean growth rate $r = 1.3 \pm 0.1\text{ years}^{-1}$ and  delay  $\tau = 3\pm1$ years across languages (see Methods and Figure \ref{fig:sup_hopf}). 

The question arises as to the nature of the words that align with the Hopf bifurcation. We found no significant contribution of words from any specific community or semantic domain. We also performed Granger tests at word level and found only a slightly higher proportion of $W\rightarrow E$ words (23\% compared to the global 22\%, $t(20) = 7.0$, $p < 0.001$). These results suggest that words are tuned to the bifurcation regardless of their interaction with economic activity.

\section*{\large Discussion}

In the first part of this work, we revisited previous findings that identified cyclic components in word usage via wavelet analysis \cite{Montemurro2001, pintos2022cognitive}. By combining Singular Spectrum Analysis (SSA) for detrending with Linear Predictive Coding (LPC) for spectral estimation, we reproduced these results with improved resolution, revealing finer spectral structure across language and databases.

Massive linguistic databases such as Google Books have been widely used to uncover important aspects of language dynamics \cite{Michel2010, Newberry2017, Amato2018}, but have also been criticized for potential genre-related biases \cite{Pechenick2015}. The main proposed strategy to mitigate this issue is to incorporate multiple languages and diverse corpora to the analyses \cite{Younes2019}. Here, we apply both controls. First, we show that the spectral  properties of word usage series are consistent across languages sourced from Google Books. Second, we compare these results with those from the curated COHA database. Despite differences in composition, both COHA and Google Books exhibit similar spectral structures for word usage. In addition to shared spectral power centered around a 12-year period, several secondary peaks and troughs are also aligned, although they appear less pronounced in COHA due to its globally shallower spectrum. Combined with the observed cross-linguistic similarities, this consistency supports the robustness of the identified linguistic dynamics.

\noindent{\bf Word usage and economic activity.} Words that oscillate in synchrony form communities that reflect semantic relationships such as is-a, has-a, synonymy, and antonymy \cite{budel2023topological, sigman2002global}. Community sizes follow a power-law distribution with large communities that closely align with broad semantic fields including government, law, economy, military, culture, government, science, and emotions. Others refer to nature, body parts, and more specific fields such as mechanics, medicine, music, geography and agriculture.

Although an objective determination of basic semantic domains is not possible \cite{wilkins1668essay, borges1942idioma}, most frameworks identify between 9 and 12 broad semantic fields, often further divided into more specific domains or subdomains \cite{moe_semantic_domains, wright2005wordnet_review}. Based on these works, we identified a set of semantic domains that encompass all the communities across languages. We found that communities related to nature, objects, the human body, domestic life, emotion, and religion show no significant relationship with GDP dynamics. In contrast, domains more closely tied to economic processes are associated with GDP variation. Specifically, we found that GDP tends to predict the dynamics of social terms, while bidirectional feedback emerges between GDP and terms related to economics and law. Government related terms show a stronger association, exhibiting both GDP driven and reciprocal dynamics. This is consistent with evidence that rising national incomes often lead to increased government spending and regulatory activity, a phenomenon known as Wagner’s Law \cite{tang2018growth}. In this sense, economic activity actively shapes government decisions.

Notably, a very strong association is observed between GDP and science-related terms, which consistently precede GDP variations. This aligns with studies suggesting that scientific knowledge precedes economic activity. For example, research on technological complexity has demonstrated that higher levels of scientific output, especially in basic science, predict economic growth over the long term \cite{hidalgo2007science, hidalgo2009economic}. Similarly, Jaffe et al. (2013) found that scientific productivity is linked to future economic growth, particularly in developing economies where investment in basic science drives long-term economic development \cite{jaffe2013productivity}. 
Likewise, military expenditures and wartime R\&D have been shown to stimulate subsequent economic growth through medical, technological and infrastructure spillovers \cite{gross2023us, benoit1973defense}.

The identification of word communities based on the synchrony of usage cycles and their comparison with global GDP, a central indicator of macroeconomic performance, yields results that are consistent with a wide range of findings from independent data sources. We note that our analyses focus exclusively on stable nouns observed consistently over the past two centuries; newly emerging terms such as those associated with technological innovation are excluded. As a result, the communities reflect core vocabularies whose associations have persisted over time. While this is a constraint, it also provides a unique opportunity to uncover strong and enduring interactions between language and real economic activity, isolated from transient lexical trends.

\noindent{\bf Dynamical criticality.} Business cycles are characterized by an upswing phase, when excess demand over supply cannot be met by the existing fixed capital. As a result, new capital assets must be created through increased investment. When demand begins to decline, the impact on output is delayed, even if the initial growth in output was driven by the increased use of existing capital \cite{Schumpeter1939}. Dynamical models that include such delays reproduce sustained oscillations through Hopf bifurcations \cite{kaleki1935macrodynamic, kurkina2017mathematical, cai2005hopf}. For example, Ercolani et al. \cite{Ercolani2014} fit a delay model to UK capital investment data and identified a statistically significant 9.6 year cycle, driven by gestation lags of about 2.4 years, which is close to the delay range around 3 years obtained from fitting our model to word usage data.

Our model for word usage \cite{pintos2022cognitive} also exhibits a Hopf bifurcation. In this case, the delay is interpreted as a characteristic time of collective attention  \cite{pintos2022cognitive}. Notably, parameter fitting indicates that oscillations operate near the bifurcation, a boundary between damped and sustained regimes. Such systems display high sensitivity and sharp frequency selectivity near resonance \cite{Mora2011}: when stimulated at their resonant frequency, the gain increases steeply, amplifying weak signals near resonance and decreasing rapidly away from it. Several biological systems take advantage of this property \cite{Mora2011}; in the mammalian auditory system, for instance, sound acts as an external input that triggers the response of cochlear cells, achieving heightened sensitivity and precise tuning by operating near a Hopf bifurcation \cite{Robles2001, Magnasco2003, Eguiluz2000}.
Word usage, although similarly poised near a Hopf bifurcation, does not exhibit a  reactive relationship with economic activity. Instead, words and economic activity influence each other over time. 
This form of tuning near a Hopf bifurcation, 
has not yet been examined from a dynamical perspective and constitutes the main objective of our ongoing investigation.

This framework may also contribute to the longstanding debate over whether economic cycles are primarily driven by external shocks 
\cite{Stadler1994}, or arise from endogenous deterministic mechanisms
\cite{Skott2023, Freeman1997}. Notably, systems near criticality are sensitive enough to exhibit both externally triggered and self-sustained oscillations. We propose that language usage, operating near a Hopf bifurcation, may simultaneously reflect sensitivity to perturbations and reveal internally generated cyclic behavior.

Our analyses support the idea that the origin of these cycles may be rooted in collective attention. The delay in word usage dynamics could reflect a cognitive process, where language shifts in response to novelty and changing social focus. In this view, cycles in language are driven by the  collective focus on new information and ideas. Economic activity follows these cognitive rhythms, with market responses such as supply-demand adjustments becoming material consequences of the novelty-driven cycles in language and culture. This framework suggests that economic cycles could ultimately be a reflection of human attention, dynamically regulated by the cognitive processes that govern language use.

\section*{\large Materials and Methods}
\label{sec:Methods}

\noindent{\bf Linguistic data.} We extracted tokens from the Corpus of Historical American English (COHA) and from the German, Spanish, French, Italian, American, and British English datasets in Google Books. The Google Books database includes digitized texts from over 25 million books (approximately 10\% of all books ever printed) published between 1500 and 2019 \cite{Michel2010}. We also collected the most stable nouns from the COHA dataset, containing over 475 million words spanning 1820 to 2019, balanced across genres (fiction, non-fiction, magazines, newspapers), facilitating standardized historical linguistic analyses \cite{davies2010corpus}.
To ensure consistent temporal coverage, we included in our analysis the fiction, non-fiction, and magazines genres from COHA, which span the full 200-year period under study. 

To warrant statistical robustness, we extracted nouns appearing in at least 180 years along the last two centuries in the COHA database, then converted them to the singular form \cite{Bird2009}. These nouns were then collected from the American and British datasets. For the other languages, we also required at least $10^6$ occurrences within the analyzed period. This left us with a core vocabulary of 4,958 nouns from COHA, Google Books American and Google Books British; 5,336 nouns from Google Books French; 3,405 nouns from Google Books Spanish; 5,987 nouns from Google Books German and 3,155 nouns from Google Books Italian.

\noindent{\bf Economic data.} We use the 2023 release of GDP data from the Maddison Project Database \cite{maddison2023}. We aggregated the GDP from the Western countries with the highest average GDP since 1820 (United States, United Kingdom, France, Germany, Italy, and Spain) into a single series with yearly samples. These countries include the majority of native speakers from the analyzed languages. All analyses are performed on the variations of the global GDP, computed as $\text{GDP}(t+1)-\text{GDP}(t)$.

\noindent{\bf Data processing.} Following standard procedures \cite{Montemurro2016, pintos2022cognitive}, raw word-usage time series were calculated for each word $j$ as $X_j(t) = N_j(t)/N(t)$, where $N_j(t)$ is the count of occurrences in year $t$, and $N(t)$ is the total word count for the corresponding language and year.

Long-term trends were extracted via Singular Spectrum Analysis (SSA) \cite{Zhigljavsky2011}, implemented using the Pyts library \cite{pyts}. SSA is a flexible, data-driven approach requiring no predetermined cutoff frequency or explicit model assumptions \cite{golyandina2013singular}. Subtracting these trends from $X(t)$ yielded small-amplitude fluctuations in word usage, $x(t)$. 

Finally, we computed the spectrum of concatenated fluctuations $x(t)$ for all nouns in each language using Linear Predictive Coding (LPC) (Figure \ref{fig:time_series}c). Following common practice \cite{rabiner1978digital,kay1988modern}, we chose an LPC order of 50, equal to one-quarter of the number of data points in each time series, sufficient to accurately resolve the prominent spectral peaks.

This combined method of SSA and LPC offers improved spectral resolution compared to previous wavelet-based analyses or approaches requiring externally defined cutoff parameters \cite{Montemurro2001, pintos2022cognitive}, resulting in more accurate identification of spectral peaks.  
After spectral analysis,  $x(t)$  series were low-pass filtered with a cutoff frequecy of $1/6$ years$^{-1}$\cite{pintos2022cognitive}.

The literature reports corrections to the global GDP mainly due to strong economic perturbations suffered during the global conflicts (\cite{korotayev2015}). Here we compared the LPC spectra generated with both series and found no significant differences: in both cases, the main peaks at 9 and 12 years, as well as the secondary ones, appeared consistently.

\noindent\textbf{Community Detection.}
Word communities were identified using the Louvain algorithm \cite{blondel2008fast}, with phase coherence between words $i$ and $j$ as the clustering measure, defined as $\frac{1}{2}\left\langle \left|e^{i \varphi_i(t)} + e^{i \varphi_j(t)} \right| \right\rangle$. The algorithm includes a tunable resolution parameter $\gamma$, which controls community size and number. We calibrated $\gamma$ by optimizing detection of known semantic shifts documented in the Oxford English Dictionary \cite{oed}, tracking communities over 75-year sliding windows with 60-year overlap (see Figure~\ref{fig:communities_time}). A value of $\gamma = 1.2$ yielded internally coherent, well-separated communities that effectively captured the semantic shifts listed in Table~\ref{table:meanings}.

Using this calibrated resolution, we clustered word communities across languages over the entire two-century period, retaining only those with more than 20 words to ensure meaningful semantic characterization. In the case of COHA, however, fewer than 6 communities met this criteria, so we selected the first 10. This process resulted in 10 communities for American English in COHA, 18 for Google American, 22 for Google British, 28 for Google German, 30 for Google French, 20 for Google Spanish, and 19 for Google Italian.

The resulting size distributions followed a power-law with high goodness of fit ($R^2 > 0.94$) across all languages, with exponents ranging from $-0.67$ to $-0.78$. An exception was COHA, which exhibited a shallower exponent of approximately $-0.56$.

\noindent\textbf{Semantic content of word communities.}
To characterize the semantic content of each word community, we used ChatGPT 4o. For each community, we submitted the full list of words and prompted the model to identify overarching semantic themes or domains, if present. This approach provided concise summaries of the dominant meanings represented in each group. In ambiguous cases, multiple prompts were used for consistency. The main words of the largest communities for each language, and their corresponding semantic domains are available in Table (\ref{table:communities}).

\noindent{\bf Directionality between word usage and GDP.} 
We are interested in identifying leading and lagging dynamics between economic activity, as reflected in GDP, and a word community, characterized by averaging and z-scoring the time series of its constituent words.
We use the Granger causality test \cite{granger1969investigating}, which evaluates whether past values of a series $X_t$ improve predictions of another series $Y_t$, beyond what is explained by $Y_t$'s own history. Formally, if the prediction error $\epsilon'$ in the expression $Y_t = \sum_{i=1}^{k} \alpha_i Y_{t-i} + \sum_{i=1}^{k} \beta_i X_{t-i} + \epsilon'_t$ is significantly lower than the error using only the first sum, then $X_t$ is said to Granger-cause $Y_t$ \cite{seabold2010statsmodels}.
For each lag $k$, the test yields one of four possible outcomes:  the GDP Granger-causes the  community ($E\rightarrow W$), the community Granger-causes GDP ($W\rightarrow E$), both directions are significant ($W\leftrightarrow E$), or neither direction is significant). 

Tests were run for lags $k$ between 20 and 40 years (roughly 1.5 to 3 characteristic periods; see Figure~\ref{fig:lag_granger}), and significance was evaluated using an F-test with a threshold of $p < 0.05$. To determine whether a semantic domain shows a preferred direction, we hypothesized that such a direction occurs more often than expected by chance, and performed a one-tailed hypergeometric test for each lag ($p < 0.05$).

To test whether words tuned at the Hopf bifurcation exhibit a preferential directionality, we compare the proportion of words associated with each Granger outcome as a function of lag, combining all languages. For each outcome, we perform a paired \textit{t}-test between the proportion in Hopf and in the rest of the corpus. The mean GDP-driving proportion was higher in Hopf (0.23) than in the rest (0.22), $t(20) = 7.0$, $p < 0.001$. Conversely, the proportions of GDP-driven and bilateral outcomes were lower in Hopf (0.20 and 0.13) compared to the rest (0.22 and 0.14), $t(20) = -7.2$, $p < 0.001$ and $t(20) = -13.7$, $p < 0.001$, respectively.

\noindent{\bf Model.} In previous work \cite{pintos2022cognitive} we modeled fluctuations in word usage $x(t)$ using the logistic equation with distributed delay \cite{Ladas1994}:

\begin{equation} \dot{x}(t)=r x(t)\left[1-\frac{1}{x^*}\int_{-\infty}^{t} G(t-t')\,x(t')\,\,dt'\right], \label{eq:lotkavolterra} \end{equation}

\noindent where $r$ is the usage growth rate, and $G(t)$ is a weighting function specifying the inhibitory influence of past values. We choose the strong kernel  $G(t)=4t/\tau^2\,e^{-2t/\tau}$, which imposes a preferred average delay $\tau$. The equilibrium $x^*$ corresponds to the long-term trend in word usage. This system exhibits a Hopf bifurcation at $\tau=4/r$ (Fig. \ref{fig:hopf}a, black curve). 

\noindent{\bf Parameter fitting.} 
Following \cite{pintos2022cognitive}, we Following \cite{pintos2022cognitive}, we integrated Equation~\ref{eq:lotkavolterra} over the grid $0 \leq r \leq 3$ and $0.1 \leq \tau \leq 12$ with step sizes of $0.01$ and $0.1$, respectively. The best fits were selected by minimizing $z(C) + 2z(D)$, where $C$ and $D$ represent the inverse correlation and the ratio between the standard deviations of the experimental and simulated data, respectively. We defined a neighborhood around the Hopf curve $\tau = 4/r$, including all cases falling between $\tau \geq 3/5$ and $\tau \leq 5/r$. The distribution of fits for each dataset along with the Hopf vicinity is shown in Figure~\ref{fig:sup_hopf}. Within this neighborhood, 71\% of the best-fitting cases fell within the Hopf region for COHA, 86\% for Google American, 91\% for Google British, 84\% for Google German, 84\% for Google Italian, 84\% for Google French, and 80\% for Google Spanish.


\noindent{\bf Data availability.}
Google Books data are publicly available at \cite{googlebooks}. Access to the COHA can be purchased directly at \cite{davies2010coha}. 
The Matlab codes that generate the figures of this work are available at \cite{repo}.

\vspace{1cm}
\begin{acknowledgments}
This research was partially funded by the Universidad de Buenos Aires (UBA) through Grant UBACyT, 20020220100181BA, the Consejo Nacional de Investigaciones Cientíﬁcas y Técnicas (CONICET) through Grant No. PIP-11220200102083CO, and the Agencia Nacional de Promoción de la Investigación, el Desarrollo Tecnológico y la Innovación through Grant No.
PICT-2020-SERIEA-00966. APP was funded by a doctoral scholarship from CONICET.
\end{acknowledgments}

\bibliography{MyCollection}

\begin{thebibliography}{57}%
\makeatletter
\providecommand \@ifxundefined [1]{%
 \@ifx{#1\undefined}
}%
\providecommand \@ifnum [1]{%
 \ifnum #1\expandafter \@firstoftwo
 \else \expandafter \@secondoftwo
 \fi
}%
\providecommand \@ifx [1]{%
 \ifx #1\expandafter \@firstoftwo
 \else \expandafter \@secondoftwo
 \fi
}%
\providecommand \natexlab [1]{#1}%
\providecommand \enquote  [1]{``#1''}%
\providecommand \bibnamefont  [1]{#1}%
\providecommand \bibfnamefont [1]{#1}%
\providecommand \citenamefont [1]{#1}%
\providecommand \href@noop [0]{\@secondoftwo}%
\providecommand \href [0]{\begingroup \@sanitize@url \@href}%
\providecommand \@href[1]{\@@startlink{#1}\@@href}%
\providecommand \@@href[1]{\endgroup#1\@@endlink}%
\providecommand \@sanitize@url [0]{\catcode `\\12\catcode `\$12\catcode `\&12\catcode `\#12\catcode `\^12\catcode `\_12\catcode `\%12\relax}%
\providecommand \@@startlink[1]{}%
\providecommand \@@endlink[0]{}%
\providecommand \url  [0]{\begingroup\@sanitize@url \@url }%
\providecommand \@url [1]{\endgroup\@href {#1}{\urlprefix }}%
\providecommand \urlprefix  [0]{URL }%
\providecommand \Eprint [0]{\href }%
\providecommand \doibase [0]{https://doi.org/}%
\providecommand \selectlanguage [0]{\@gobble}%
\providecommand \bibinfo  [0]{\@secondoftwo}%
\providecommand \bibfield  [0]{\@secondoftwo}%
\providecommand \translation [1]{[#1]}%
\providecommand \BibitemOpen [0]{}%
\providecommand \bibitemStop [0]{}%
\providecommand \bibitemNoStop [0]{.\EOS\space}%
\providecommand \EOS [0]{\spacefactor3000\relax}%
\providecommand \BibitemShut  [1]{\csname bibitem#1\endcsname}%
\let\auto@bib@innerbib\@empty
\bibitem [{\citenamefont {Zipf}(1933)}]{Zipf1933}%
  \BibitemOpen
  \bibfield  {author} {\bibinfo {author} {\bibfnamefont {G.~K.}\ \bibnamefont {Zipf}},\ }\href {https://doi.org/10.4159/harvard.9780674434929} {\bibfield  {journal} {\bibinfo  {journal} {Language}\ }\textbf {\bibinfo {volume} {9}},\ \bibinfo {pages} {89} (\bibinfo {year} {1933})}\BibitemShut {NoStop}%
\bibitem [{\citenamefont {Aitchison}\ \emph {et~al.}(2016)\citenamefont {Aitchison}, \citenamefont {Corradi},\ and\ \citenamefont {Latham}}]{aitchison2016zipf}%
  \BibitemOpen
  \bibfield  {author} {\bibinfo {author} {\bibfnamefont {L.}~\bibnamefont {Aitchison}}, \bibinfo {author} {\bibfnamefont {N.}~\bibnamefont {Corradi}},\ and\ \bibinfo {author} {\bibfnamefont {P.~E.}\ \bibnamefont {Latham}},\ }\href@noop {} {\bibfield  {journal} {\bibinfo  {journal} {PLoS computational biology}\ }\textbf {\bibinfo {volume} {12}},\ \bibinfo {pages} {e1005110} (\bibinfo {year} {2016})}\BibitemShut {NoStop}%
\bibitem [{\citenamefont {Montemurro}\ and\ \citenamefont {Zanette}(2016)}]{Montemurro2016}%
  \BibitemOpen
  \bibfield  {author} {\bibinfo {author} {\bibfnamefont {M.~A.}\ \bibnamefont {Montemurro}}\ and\ \bibinfo {author} {\bibfnamefont {D.~H.}\ \bibnamefont {Zanette}},\ }\href {https://doi.org/10.1057/palcomms.2016.84} {\bibfield  {journal} {\bibinfo  {journal} {Palgrave Communications}\ }\textbf {\bibinfo {volume} {2}},\ \bibinfo {pages} {1} (\bibinfo {year} {2016})}\BibitemShut {NoStop}%
\bibitem [{\citenamefont {Pintos}\ \emph {et~al.}(2022)\citenamefont {Pintos}, \citenamefont {Shalom}, \citenamefont {Tagliazucchi}, \citenamefont {Mindlin},\ and\ \citenamefont {Trevisan}}]{pintos2022cognitive}%
  \BibitemOpen
  \bibfield  {author} {\bibinfo {author} {\bibfnamefont {A.~P.}\ \bibnamefont {Pintos}}, \bibinfo {author} {\bibfnamefont {D.~E.}\ \bibnamefont {Shalom}}, \bibinfo {author} {\bibfnamefont {E.}~\bibnamefont {Tagliazucchi}}, \bibinfo {author} {\bibfnamefont {G.}~\bibnamefont {Mindlin}},\ and\ \bibinfo {author} {\bibfnamefont {M.}~\bibnamefont {Trevisan}},\ }\href@noop {} {\bibfield  {journal} {\bibinfo  {journal} {Chaos, Solitons \& Fractals}\ }\textbf {\bibinfo {volume} {161}},\ \bibinfo {pages} {112327} (\bibinfo {year} {2022})}\BibitemShut {NoStop}%
\bibitem [{\citenamefont {Michel}\ \emph {et~al.}(2010)\citenamefont {Michel}, \citenamefont {Shen}, \citenamefont {Aiden}, \citenamefont {Veres}, \citenamefont {Gray}, \citenamefont {Pickett}, \citenamefont {Hoiberg}, \citenamefont {Clancy}, \citenamefont {Norvig}, \citenamefont {Orwant}, \citenamefont {Pinker}, \citenamefont {Nowak},\ and\ \citenamefont {Aiden}}]{Michel2010}%
  \BibitemOpen
  \bibfield  {author} {\bibinfo {author} {\bibfnamefont {J.-B.}\ \bibnamefont {Michel}}, \bibinfo {author} {\bibfnamefont {Y.~K.}\ \bibnamefont {Shen}}, \bibinfo {author} {\bibfnamefont {a.~P.}\ \bibnamefont {Aiden}}, \bibinfo {author} {\bibfnamefont {A.}~\bibnamefont {Veres}}, \bibinfo {author} {\bibfnamefont {M.~K.}\ \bibnamefont {Gray}}, \bibinfo {author} {\bibfnamefont {J.~P.}\ \bibnamefont {Pickett}}, \bibinfo {author} {\bibfnamefont {D.}~\bibnamefont {Hoiberg}}, \bibinfo {author} {\bibfnamefont {D.}~\bibnamefont {Clancy}}, \bibinfo {author} {\bibfnamefont {P.}~\bibnamefont {Norvig}}, \bibinfo {author} {\bibfnamefont {J.}~\bibnamefont {Orwant}}, \bibinfo {author} {\bibfnamefont {S.}~\bibnamefont {Pinker}}, \bibinfo {author} {\bibfnamefont {M.~a.}\ \bibnamefont {Nowak}},\ and\ \bibinfo {author} {\bibfnamefont {E.~L.}\ \bibnamefont {Aiden}},\ }\href {https://doi.org/10.1126/science.1199644} {\bibfield  {journal} {\bibinfo  {journal} {Science}\ }\textbf {\bibinfo {volume} {331}},\ \bibinfo {pages} {176}
  (\bibinfo {year} {2010})}\BibitemShut {NoStop}%
\bibitem [{\citenamefont {Davies}(2010{\natexlab{a}})}]{davies2010corpus}%
  \BibitemOpen
  \bibfield  {author} {\bibinfo {author} {\bibfnamefont {M.}~\bibnamefont {Davies}},\ }\href@noop {} {\emph {\bibinfo {title} {The corpus of historical American English: COHA}}}\ (\bibinfo  {publisher} {BYE, Brigham Young University},\ \bibinfo {year} {2010})\BibitemShut {NoStop}%
\bibitem [{\citenamefont {Pechenick}\ \emph {et~al.}(2015)\citenamefont {Pechenick}, \citenamefont {Danforth},\ and\ \citenamefont {Dodds}}]{Pechenick2015}%
  \BibitemOpen
  \bibfield  {author} {\bibinfo {author} {\bibfnamefont {E.~A.}\ \bibnamefont {Pechenick}}, \bibinfo {author} {\bibfnamefont {C.~M.}\ \bibnamefont {Danforth}},\ and\ \bibinfo {author} {\bibfnamefont {P.~S.}\ \bibnamefont {Dodds}},\ }\href {https://doi.org/10.1371/journal.pone.0137041} {\bibfield  {journal} {\bibinfo  {journal} {PLoS ONE}\ }\textbf {\bibinfo {volume} {10}},\ \bibinfo {pages} {1} (\bibinfo {year} {2015})},\ \Eprint {https://arxiv.org/abs/1501.00960} {arXiv:1501.00960} \BibitemShut {NoStop}%
\bibitem [{\citenamefont {Younes}\ and\ \citenamefont {Reips}(2019)}]{Younes2019}%
  \BibitemOpen
  \bibfield  {author} {\bibinfo {author} {\bibfnamefont {N.}~\bibnamefont {Younes}}\ and\ \bibinfo {author} {\bibfnamefont {U.~D.}\ \bibnamefont {Reips}},\ }\href {https://doi.org/10.1371/journal.pone.0213554} {\bibfield  {journal} {\bibinfo  {journal} {PLoS ONE}\ }\textbf {\bibinfo {volume} {14}},\ \bibinfo {pages} {1} (\bibinfo {year} {2019})}\BibitemShut {NoStop}%
\bibitem [{\citenamefont {Lansdall-Welfare}\ \emph {et~al.}(2017)\citenamefont {Lansdall-Welfare}, \citenamefont {Sudhahar}, \citenamefont {Thompson}, \citenamefont {Lewis}, \citenamefont {Team},\ and\ \citenamefont {Cristianini}}]{lansdall2017content}%
  \BibitemOpen
  \bibfield  {author} {\bibinfo {author} {\bibfnamefont {T.}~\bibnamefont {Lansdall-Welfare}}, \bibinfo {author} {\bibfnamefont {S.}~\bibnamefont {Sudhahar}}, \bibinfo {author} {\bibfnamefont {J.}~\bibnamefont {Thompson}}, \bibinfo {author} {\bibfnamefont {J.}~\bibnamefont {Lewis}}, \bibinfo {author} {\bibfnamefont {F.~N.}\ \bibnamefont {Team}},\ and\ \bibinfo {author} {\bibfnamefont {N.}~\bibnamefont {Cristianini}},\ }\href@noop {} {\bibfield  {journal} {\bibinfo  {journal} {Proceedings of the National Academy of Sciences}\ }\textbf {\bibinfo {volume} {114}},\ \bibinfo {pages} {E457} (\bibinfo {year} {2017})}\BibitemShut {NoStop}%
\bibitem [{\citenamefont {Juglar}(1862)}]{juglar1862}%
  \BibitemOpen
  \bibfield  {author} {\bibinfo {author} {\bibfnamefont {C.}~\bibnamefont {Juglar}},\ }\href@noop {} {\emph {\bibinfo {title} {Des Crises Commerciales et de Leur Retour Périodique}}}\ (\bibinfo  {publisher} {Guillaumin},\ \bibinfo {address} {Paris},\ \bibinfo {year} {1862})\BibitemShut {NoStop}%
\bibitem [{\citenamefont {Kurkina}(2017)}]{kurkina2017mathematical}%
  \BibitemOpen
  \bibfield  {author} {\bibinfo {author} {\bibfnamefont {E.~S.}\ \bibnamefont {Kurkina}},\ }\href@noop {} {\bibfield  {journal} {\bibinfo  {journal} {Computational Mathematics and Modeling}\ }\textbf {\bibinfo {volume} {28}},\ \bibinfo {pages} {377} (\bibinfo {year} {2017})}\BibitemShut {NoStop}%
\bibitem [{\citenamefont {Schumpeter}(1939)}]{Schumpeter1939}%
  \BibitemOpen
  \bibfield  {author} {\bibinfo {author} {\bibfnamefont {J.~A.}\ \bibnamefont {Schumpeter}},\ }\href {https://archive.org/details/businesscycles0001unse} {\emph {\bibinfo {title} {Business Cycles: A Theoretical, Historical, and Statistical Analysis of the Capitalist Process}}}\ (\bibinfo  {publisher} {McGraw-Hill Book Company},\ \bibinfo {address} {New York and London},\ \bibinfo {year} {1939})\BibitemShut {NoStop}%
\bibitem [{\citenamefont {Kaleki}(1935)}]{kaleki1935macrodynamic}%
  \BibitemOpen
  \bibfield  {author} {\bibinfo {author} {\bibfnamefont {M.}~\bibnamefont {Kaleki}},\ }\href@noop {} {\bibfield  {journal} {\bibinfo  {journal} {Econometrica}\ }\textbf {\bibinfo {volume} {3}},\ \bibinfo {pages} {327} (\bibinfo {year} {1935})}\BibitemShut {NoStop}%
\bibitem [{\citenamefont {Cai}(2005)}]{cai2005hopf}%
  \BibitemOpen
  \bibfield  {author} {\bibinfo {author} {\bibfnamefont {J.}~\bibnamefont {Cai}},\ }\href@noop {} {\bibfield  {journal} {\bibinfo  {journal} {Electronic Journal of Differential Equations (EJDE)[electronic only]}\ }\textbf {\bibinfo {volume} {2005}},\ \bibinfo {pages} {Paper} (\bibinfo {year} {2005})}\BibitemShut {NoStop}%
\bibitem [{\citenamefont {Kitchin}(1923)}]{kitchin1923}%
  \BibitemOpen
  \bibfield  {author} {\bibinfo {author} {\bibfnamefont {J.}~\bibnamefont {Kitchin}},\ }\href {https://doi.org/10.2307/1927031} {\bibfield  {journal} {\bibinfo  {journal} {The Review of Economics and Statistics}\ }\textbf {\bibinfo {volume} {5}},\ \bibinfo {pages} {10} (\bibinfo {year} {1923})}\BibitemShut {NoStop}%
\bibitem [{\citenamefont {Kuznets}(1958)}]{kuznets1958quantitative}%
  \BibitemOpen
  \bibfield  {author} {\bibinfo {author} {\bibfnamefont {S.}~\bibnamefont {Kuznets}},\ }\href@noop {} {\bibfield  {journal} {\bibinfo  {journal} {Economic Development and Cultural Change}\ }\textbf {\bibinfo {volume} {6}},\ \bibinfo {pages} {1} (\bibinfo {year} {1958})}\BibitemShut {NoStop}%
\bibitem [{\citenamefont {Korotayev}\ and\ \citenamefont {Tsirel}(2010)}]{korotayev2010spectral}%
  \BibitemOpen
  \bibfield  {author} {\bibinfo {author} {\bibfnamefont {A.~V.}\ \bibnamefont {Korotayev}}\ and\ \bibinfo {author} {\bibfnamefont {S.~V.}\ \bibnamefont {Tsirel}},\ }\href@noop {} {\bibfield  {journal} {\bibinfo  {journal} {Structure and Dynamics}\ }\textbf {\bibinfo {volume} {4}} (\bibinfo {year} {2010})}\BibitemShut {NoStop}%
\bibitem [{\citenamefont {Montemurro}(2001)}]{Montemurro2001}%
  \BibitemOpen
  \bibfield  {author} {\bibinfo {author} {\bibfnamefont {M.~A.}\ \bibnamefont {Montemurro}},\ }\href {https://doi.org/10.1016/S0378-4371(01)00355-7} {\bibfield  {journal} {\bibinfo  {journal} {Physica A: Statistical Mechanics and its Applications}\ }\textbf {\bibinfo {volume} {300}},\ \bibinfo {pages} {567} (\bibinfo {year} {2001})},\ \Eprint {https://arxiv.org/abs/0104066} {arXiv:0104066 [cond-mat]} \BibitemShut {NoStop}%
\bibitem [{\citenamefont {Petersen}\ \emph {et~al.}(2012)\citenamefont {Petersen}, \citenamefont {Tenenbaum}, \citenamefont {Havlin}, \citenamefont {Stanley},\ and\ \citenamefont {Perc}}]{petersen2012languages}%
  \BibitemOpen
  \bibfield  {author} {\bibinfo {author} {\bibfnamefont {A.~M.}\ \bibnamefont {Petersen}}, \bibinfo {author} {\bibfnamefont {J.~N.}\ \bibnamefont {Tenenbaum}}, \bibinfo {author} {\bibfnamefont {S.}~\bibnamefont {Havlin}}, \bibinfo {author} {\bibfnamefont {H.~E.}\ \bibnamefont {Stanley}},\ and\ \bibinfo {author} {\bibfnamefont {M.}~\bibnamefont {Perc}},\ }\href@noop {} {\bibfield  {journal} {\bibinfo  {journal} {Scientific reports}\ }\textbf {\bibinfo {volume} {2}},\ \bibinfo {pages} {943} (\bibinfo {year} {2012})}\BibitemShut {NoStop}%
\bibitem [{\citenamefont {Newberry}\ \emph {et~al.}(2017)\citenamefont {Newberry}, \citenamefont {Ahern}, \citenamefont {Clark},\ and\ \citenamefont {Plotkin}}]{Newberry2017}%
  \BibitemOpen
  \bibfield  {author} {\bibinfo {author} {\bibfnamefont {M.~G.}\ \bibnamefont {Newberry}}, \bibinfo {author} {\bibfnamefont {C.~A.}\ \bibnamefont {Ahern}}, \bibinfo {author} {\bibfnamefont {R.}~\bibnamefont {Clark}},\ and\ \bibinfo {author} {\bibfnamefont {J.~B.}\ \bibnamefont {Plotkin}},\ }\href {https://doi.org/10.1038/nature24455} {\bibfield  {journal} {\bibinfo  {journal} {Nature Publishing Group}\ }\textbf {\bibinfo {volume} {551}},\ \bibinfo {pages} {223} (\bibinfo {year} {2017})}\BibitemShut {NoStop}%
\bibitem [{\citenamefont {Amato}\ \emph {et~al.}(2018)\citenamefont {Amato}, \citenamefont {Lacasa}, \citenamefont {D{\'\i}az-Guilera},\ and\ \citenamefont {Baronchelli}}]{Amato2018}%
  \BibitemOpen
  \bibfield  {author} {\bibinfo {author} {\bibfnamefont {R.}~\bibnamefont {Amato}}, \bibinfo {author} {\bibfnamefont {L.}~\bibnamefont {Lacasa}}, \bibinfo {author} {\bibfnamefont {A.}~\bibnamefont {D{\'\i}az-Guilera}},\ and\ \bibinfo {author} {\bibfnamefont {A.}~\bibnamefont {Baronchelli}},\ }\href@noop {} {\bibfield  {journal} {\bibinfo  {journal} {Proceedings of the National Academy of Sciences}\ }\textbf {\bibinfo {volume} {115}},\ \bibinfo {pages} {8260} (\bibinfo {year} {2018})}\BibitemShut {NoStop}%
\bibitem [{\citenamefont {Budel}\ \emph {et~al.}(2023)\citenamefont {Budel}, \citenamefont {Jin}, \citenamefont {Van~Mieghem},\ and\ \citenamefont {Kitsak}}]{budel2023topological}%
  \BibitemOpen
  \bibfield  {author} {\bibinfo {author} {\bibfnamefont {G.}~\bibnamefont {Budel}}, \bibinfo {author} {\bibfnamefont {Y.}~\bibnamefont {Jin}}, \bibinfo {author} {\bibfnamefont {P.}~\bibnamefont {Van~Mieghem}},\ and\ \bibinfo {author} {\bibfnamefont {M.}~\bibnamefont {Kitsak}},\ }\href@noop {} {\bibfield  {journal} {\bibinfo  {journal} {Scientific Reports}\ }\textbf {\bibinfo {volume} {13}},\ \bibinfo {pages} {11728} (\bibinfo {year} {2023})}\BibitemShut {NoStop}%
\bibitem [{\citenamefont {Sigman}\ and\ \citenamefont {Cecchi}(2002)}]{sigman2002global}%
  \BibitemOpen
  \bibfield  {author} {\bibinfo {author} {\bibfnamefont {M.}~\bibnamefont {Sigman}}\ and\ \bibinfo {author} {\bibfnamefont {G.~A.}\ \bibnamefont {Cecchi}},\ }\href@noop {} {\bibfield  {journal} {\bibinfo  {journal} {Proceedings of the National Academy of Sciences}\ }\textbf {\bibinfo {volume} {99}},\ \bibinfo {pages} {1742} (\bibinfo {year} {2002})}\BibitemShut {NoStop}%
\bibitem [{\citenamefont {Wilkins}(1668)}]{wilkins1668essay}%
  \BibitemOpen
  \bibfield  {author} {\bibinfo {author} {\bibfnamefont {J.}~\bibnamefont {Wilkins}},\ }\href@noop {} {\emph {\bibinfo {title} {An Essay Towards a Real Character, and a Philosophical Language}}}\ (\bibinfo  {publisher} {Samuel Smith and Benjamin Walford},\ \bibinfo {address} {London},\ \bibinfo {year} {1668})\ \bibinfo {note} {proposes a universal taxonomic language with 40 categories}\BibitemShut {NoStop}%
\bibitem [{\citenamefont {Borges}(1942)}]{borges1942idioma}%
  \BibitemOpen
  \bibfield  {author} {\bibinfo {author} {\bibfnamefont {J.~L.}\ \bibnamefont {Borges}},\ }in\ \href@noop {} {\emph {\bibinfo {booktitle} {Otras inquisiciones}}}\ (\bibinfo  {publisher} {Sur},\ \bibinfo {address} {Buenos Aires},\ \bibinfo {year} {1942})\ \bibinfo {note} {reprinted in various editions; English translation available in *Selected Non-Fictions* (Penguin, 2000)}\BibitemShut {NoStop}%
\bibitem [{\citenamefont {Moe}\ and\ \citenamefont {International}(2025)}]{moe_semantic_domains}%
  \BibitemOpen
  \bibfield  {author} {\bibinfo {author} {\bibfnamefont {R.}~\bibnamefont {Moe}}\ and\ \bibinfo {author} {\bibfnamefont {S.}~\bibnamefont {International}},\ }\href@noop {} {\bibinfo {title} {Semantic domains}},\ \bibinfo {howpublished} {{Online resource listed under nine major headings, with ca.\ 1800 subdomains}} (\bibinfo {year} {2025}),\ \bibinfo {note} {accessed via semdom.org}\BibitemShut {NoStop}%
\bibitem [{\citenamefont {Wright}\ \emph {et~al.}(2005)\citenamefont {Wright}, \citenamefont {Grishman},\ and\ \citenamefont {Aronoff}}]{wright2005wordnet_review}%
  \BibitemOpen
  \bibfield  {author} {\bibinfo {author} {\bibfnamefont {J.}~\bibnamefont {Wright}}, \bibinfo {author} {\bibfnamefont {R.}~\bibnamefont {Grishman}},\ and\ \bibinfo {author} {\bibfnamefont {M.}~\bibnamefont {Aronoff}},\ }\href@noop {} {\bibfield  {journal} {\bibinfo  {journal} {Journal of Philosophical Studies}\ } (\bibinfo {year} {2005})},\ \bibinfo {note} {defines core semantic fields such as entity, event, cognition, etc.}\BibitemShut {Stop}%
\bibitem [{\citenamefont {Tang}\ and\ \citenamefont {Tan}(2018)}]{tang2018growth}%
  \BibitemOpen
  \bibfield  {author} {\bibinfo {author} {\bibfnamefont {C.}~\bibnamefont {Tang}}\ and\ \bibinfo {author} {\bibfnamefont {E.}~\bibnamefont {Tan}},\ }\href@noop {} {\bibfield  {journal} {\bibinfo  {journal} {SAGE Open}\ }\textbf {\bibinfo {volume} {8}},\ \bibinfo {pages} {1} (\bibinfo {year} {2018})}\BibitemShut {NoStop}%
\bibitem [{\citenamefont {Hidalgo}\ and\ \citenamefont {Hausmann}(2007)}]{hidalgo2007science}%
  \BibitemOpen
  \bibfield  {author} {\bibinfo {author} {\bibfnamefont {C.}~\bibnamefont {Hidalgo}}\ and\ \bibinfo {author} {\bibfnamefont {R.}~\bibnamefont {Hausmann}},\ }\href@noop {} {\bibfield  {journal} {\bibinfo  {journal} {Science}\ }\textbf {\bibinfo {volume} {317}},\ \bibinfo {pages} {482} (\bibinfo {year} {2007})}\BibitemShut {NoStop}%
\bibitem [{\citenamefont {Hidalgo}\ and\ \citenamefont {Hausmann}(2009)}]{hidalgo2009economic}%
  \BibitemOpen
  \bibfield  {author} {\bibinfo {author} {\bibfnamefont {C.}~\bibnamefont {Hidalgo}}\ and\ \bibinfo {author} {\bibfnamefont {R.}~\bibnamefont {Hausmann}},\ }\href@noop {} {\bibfield  {journal} {\bibinfo  {journal} {PNAS}\ }\textbf {\bibinfo {volume} {106}},\ \bibinfo {pages} {10570} (\bibinfo {year} {2009})}\BibitemShut {NoStop}%
\bibitem [{\citenamefont {Jaffe}\ \emph {et~al.}(2013)\citenamefont {Jaffe}, \citenamefont {Caicedo}, \citenamefont {Manzanares}, \citenamefont {Gil}, \citenamefont {Rios}, \citenamefont {Florez}, \citenamefont {Montoreano},\ and\ \citenamefont {Davila}}]{jaffe2013productivity}%
  \BibitemOpen
  \bibfield  {author} {\bibinfo {author} {\bibfnamefont {K.}~\bibnamefont {Jaffe}}, \bibinfo {author} {\bibfnamefont {M.}~\bibnamefont {Caicedo}}, \bibinfo {author} {\bibfnamefont {M.}~\bibnamefont {Manzanares}}, \bibinfo {author} {\bibfnamefont {M.}~\bibnamefont {Gil}}, \bibinfo {author} {\bibfnamefont {A.}~\bibnamefont {Rios}}, \bibinfo {author} {\bibfnamefont {A.}~\bibnamefont {Florez}}, \bibinfo {author} {\bibfnamefont {C.}~\bibnamefont {Montoreano}},\ and\ \bibinfo {author} {\bibfnamefont {V.}~\bibnamefont {Davila}},\ }\href {https://doi.org/10.1371/journal.pone.0066239} {\bibfield  {journal} {\bibinfo  {journal} {PLOS ONE}\ }\textbf {\bibinfo {volume} {8}},\ \bibinfo {pages} {e66239} (\bibinfo {year} {2013})}\BibitemShut {NoStop}%
\bibitem [{\citenamefont {Gross}\ and\ \citenamefont {Sampat}(2023)}]{gross2023us}%
  \BibitemOpen
  \bibfield  {author} {\bibinfo {author} {\bibfnamefont {D.}~\bibnamefont {Gross}}\ and\ \bibinfo {author} {\bibfnamefont {B.}~\bibnamefont {Sampat}},\ }\href@noop {} {\bibfield  {journal} {\bibinfo  {journal} {American Economic Review}\ }\textbf {\bibinfo {volume} {113}},\ \bibinfo {pages} {3323} (\bibinfo {year} {2023})}\BibitemShut {NoStop}%
\bibitem [{\citenamefont {Benoit}(1973)}]{benoit1973defense}%
  \BibitemOpen
  \bibfield  {author} {\bibinfo {author} {\bibfnamefont {E.}~\bibnamefont {Benoit}},\ }\href@noop {} {\emph {\bibinfo {title} {Defense and Economic Growth in Developing Countries}}}\ (\bibinfo  {publisher} {Lexington Books},\ \bibinfo {year} {1973})\BibitemShut {NoStop}%
\bibitem [{\citenamefont {Ercolani}(2014)}]{Ercolani2014}%
  \BibitemOpen
  \bibfield  {author} {\bibinfo {author} {\bibfnamefont {J.~S.}\ \bibnamefont {Ercolani}},\ }\href@noop {} {\bibfield  {journal} {\bibinfo  {journal} {The Manchester School}\ }\textbf {\bibinfo {volume} {82}},\ \bibinfo {pages} {620} (\bibinfo {year} {2014})}\BibitemShut {NoStop}%
\bibitem [{\citenamefont {Mora}\ and\ \citenamefont {Bialek}(2011)}]{Mora2011}%
  \BibitemOpen
  \bibfield  {author} {\bibinfo {author} {\bibfnamefont {T.}~\bibnamefont {Mora}}\ and\ \bibinfo {author} {\bibfnamefont {W.}~\bibnamefont {Bialek}},\ }\href {https://doi.org/10.1007/s10955-011-0229-4} {\bibfield  {journal} {\bibinfo  {journal} {Journal of Statistical Physics}\ }\textbf {\bibinfo {volume} {144}},\ \bibinfo {pages} {268} (\bibinfo {year} {2011})},\ \Eprint {https://arxiv.org/abs/1012.2242} {arXiv:1012.2242} \BibitemShut {NoStop}%
\bibitem [{\citenamefont {Robles}\ and\ \citenamefont {Ruggero}(2001)}]{Robles2001}%
  \BibitemOpen
  \bibfield  {author} {\bibinfo {author} {\bibfnamefont {L.}~\bibnamefont {Robles}}\ and\ \bibinfo {author} {\bibfnamefont {M.~A.}\ \bibnamefont {Ruggero}},\ }\href {https://doi.org/10.1152/physrev.2001.81.3.1305} {\bibfield  {journal} {\bibinfo  {journal} {Physiological Reviews}\ }\textbf {\bibinfo {volume} {81}},\ \bibinfo {pages} {1305} (\bibinfo {year} {2001})}\BibitemShut {NoStop}%
\bibitem [{\citenamefont {Magnasco}(2003)}]{Magnasco2003}%
  \BibitemOpen
  \bibfield  {author} {\bibinfo {author} {\bibfnamefont {M.~O.}\ \bibnamefont {Magnasco}},\ }\href {https://doi.org/10.1103/PhysRevLett.90.058101} {\bibfield  {journal} {\bibinfo  {journal} {Physical review letters}\ }\textbf {\bibinfo {volume} {90}},\ \bibinfo {pages} {058101} (\bibinfo {year} {2003})},\ \Eprint {https://arxiv.org/abs/0311577} {arXiv:0311577 [cond-mat]} \BibitemShut {NoStop}%
\bibitem [{\citenamefont {Egu{\'{i}}luz}\ \emph {et~al.}(2000)\citenamefont {Egu{\'{i}}luz}, \citenamefont {Ospeck}, \citenamefont {Choe}, \citenamefont {Hudspeth},\ and\ \citenamefont {Magnasco}}]{Eguiluz2000}%
  \BibitemOpen
  \bibfield  {author} {\bibinfo {author} {\bibfnamefont {V.~M.}\ \bibnamefont {Egu{\'{i}}luz}}, \bibinfo {author} {\bibfnamefont {M.}~\bibnamefont {Ospeck}}, \bibinfo {author} {\bibfnamefont {Y.}~\bibnamefont {Choe}}, \bibinfo {author} {\bibfnamefont {a.~J.}\ \bibnamefont {Hudspeth}},\ and\ \bibinfo {author} {\bibfnamefont {M.~O.}\ \bibnamefont {Magnasco}},\ }\href {https://doi.org/10.1103/PhysRevLett.84.5232} {\bibfield  {journal} {\bibinfo  {journal} {Physical review letters}\ }\textbf {\bibinfo {volume} {84}},\ \bibinfo {pages} {5232} (\bibinfo {year} {2000})},\ \Eprint {https://arxiv.org/abs/0005042} {arXiv:0005042 [nlin]} \BibitemShut {NoStop}%
\bibitem [{\citenamefont {Stadler}(1994)}]{Stadler1994}%
  \BibitemOpen
  \bibfield  {author} {\bibinfo {author} {\bibfnamefont {G.~W.}\ \bibnamefont {Stadler}},\ }\href {https://www.econ.ucdavis.edu/faculty/kdsalyer/LECTURES/Ecn200e/Stadler.pdf} {\bibfield  {journal} {\bibinfo  {journal} {Journal of Economic Literature}\ }\textbf {\bibinfo {volume} {32}},\ \bibinfo {pages} {1750} (\bibinfo {year} {1994})}\BibitemShut {NoStop}%
\bibitem [{\citenamefont {Skott}(2023)}]{Skott2023}%
  \BibitemOpen
  \bibfield  {author} {\bibinfo {author} {\bibfnamefont {P.}~\bibnamefont {Skott}},\ }\href {https://www.sciencedirect.com/science/article/pii/S0167268123001063} {\bibfield  {journal} {\bibinfo  {journal} {European Economic Review}\ } (\bibinfo {year} {2023})}\BibitemShut {NoStop}%
\bibitem [{\citenamefont {Freeman}(1997)}]{Freeman1997}%
  \BibitemOpen
  \bibfield  {author} {\bibinfo {author} {\bibfnamefont {C.}~\bibnamefont {Freeman}},\ }\href {https://carlotaperez.org/wp-content/downloads/publications/theoretical-framework/CP%20Freeman%201997%20on%20JAS%20revisited%20in%20EJESS%202015.pdf} {\bibfield  {journal} {\bibinfo  {journal} {European Journal of the History of Economic Thought}\ }\textbf {\bibinfo {volume} {4}},\ \bibinfo {pages} {263} (\bibinfo {year} {1997})}\BibitemShut {NoStop}%
\bibitem [{\citenamefont {Bird}\ \emph {et~al.}(2009)\citenamefont {Bird}, \citenamefont {Ewan},\ and\ \citenamefont {Loper}}]{Bird2009}%
  \BibitemOpen
  \bibfield  {author} {\bibinfo {author} {\bibfnamefont {S.}~\bibnamefont {Bird}}, \bibinfo {author} {\bibfnamefont {K.}~\bibnamefont {Ewan}},\ and\ \bibinfo {author} {\bibfnamefont {E.}~\bibnamefont {Loper}},\ }\href@noop {} {\emph {\bibinfo {title} {{Natural Language Processing with Python}}}}\ (\bibinfo  {publisher} {O-Reilly},\ \bibinfo {address} {Beijing Cambridge Farnham Koln Sebastopol Taipei Tokyo},\ \bibinfo {year} {2009})\BibitemShut {NoStop}%
\bibitem [{\citenamefont {Bolt}\ \emph {et~al.}(2023)\citenamefont {Bolt}, \citenamefont {van Zanden} \emph {et~al.}}]{maddison2023}%
  \BibitemOpen
  \bibfield  {author} {\bibinfo {author} {\bibfnamefont {J.}~\bibnamefont {Bolt}}, \bibinfo {author} {\bibfnamefont {J.~L.}\ \bibnamefont {van Zanden}}, \emph {et~al.},\ }\href@noop {} {\bibinfo {title} {Maddison project database, version 2023}} (\bibinfo {year} {2023}),\ \bibinfo {note} {\url{https://www.rug.nl/ggdc/historicaldevelopment/maddison/}}\BibitemShut {NoStop}%
\bibitem [{\citenamefont {Zhigljavsky}(2011)}]{Zhigljavsky2011}%
  \BibitemOpen
  \bibfield  {author} {\bibinfo {author} {\bibfnamefont {A.}~\bibnamefont {Zhigljavsky}},\ }\href {https://doi.org/10.1007/978-3-642-04898-2_521} {\emph {\bibinfo {title} {International Encyclopedia of Statistical Science}}}\ (\bibinfo  {publisher} {Springer},\ \bibinfo {year} {2011})\ pp.\ \bibinfo {pages} {1335--1337}\BibitemShut {NoStop}%
\bibitem [{\citenamefont {Faouzi}\ and\ \citenamefont {Janati}(2020)}]{pyts}%
  \BibitemOpen
  \bibfield  {author} {\bibinfo {author} {\bibfnamefont {J.}~\bibnamefont {Faouzi}}\ and\ \bibinfo {author} {\bibfnamefont {H.}~\bibnamefont {Janati}},\ }\href {http://jmlr.org/papers/v21/19-763.html} {\bibfield  {journal} {\bibinfo  {journal} {Journal of Machine Learning Research}\ }\textbf {\bibinfo {volume} {21}},\ \bibinfo {pages} {1} (\bibinfo {year} {2020})}\BibitemShut {NoStop}%
\bibitem [{\citenamefont {Golyandina}\ and\ \citenamefont {Zhigljavsky}(2013)}]{golyandina2013singular}%
  \BibitemOpen
  \bibfield  {author} {\bibinfo {author} {\bibfnamefont {N.}~\bibnamefont {Golyandina}}\ and\ \bibinfo {author} {\bibfnamefont {A.}~\bibnamefont {Zhigljavsky}},\ }\href {https://doi.org/10.1007/978-3-642-34913-3} {\emph {\bibinfo {title} {Singular Spectrum Analysis for Time Series}}}\ (\bibinfo  {publisher} {Springer},\ \bibinfo {year} {2013})\BibitemShut {NoStop}%
\bibitem [{\citenamefont {Rabiner}\ and\ \citenamefont {Schafer}(1978)}]{rabiner1978digital}%
  \BibitemOpen
  \bibfield  {author} {\bibinfo {author} {\bibfnamefont {L.~R.}\ \bibnamefont {Rabiner}}\ and\ \bibinfo {author} {\bibfnamefont {R.~W.}\ \bibnamefont {Schafer}},\ }\href@noop {} {\emph {\bibinfo {title} {Digital Processing of Speech Signals}}}\ (\bibinfo  {publisher} {Prentice-Hall},\ \bibinfo {address} {Englewood Cliffs, NJ},\ \bibinfo {year} {1978})\BibitemShut {NoStop}%
\bibitem [{\citenamefont {Kay}(1988)}]{kay1988modern}%
  \BibitemOpen
  \bibfield  {author} {\bibinfo {author} {\bibfnamefont {S.~M.}\ \bibnamefont {Kay}},\ }\href@noop {} {\emph {\bibinfo {title} {Modern Spectral Estimation: Theory and Application}}}\ (\bibinfo  {publisher} {Prentice-Hall},\ \bibinfo {address} {Englewood Cliffs, NJ},\ \bibinfo {year} {1988})\BibitemShut {NoStop}%
\bibitem [{\citenamefont {Korotayev}\ and\ \citenamefont {Tsirel}(2015)}]{korotayev2015}%
  \BibitemOpen
  \bibfield  {author} {\bibinfo {author} {\bibfnamefont {A.~V.}\ \bibnamefont {Korotayev}}\ and\ \bibinfo {author} {\bibfnamefont {S.~V.}\ \bibnamefont {Tsirel}},\ }\href {https://doi.org/10.1111/sapm.12074} {\bibfield  {journal} {\bibinfo  {journal} {Studies in Applied Mathematics}\ }\textbf {\bibinfo {volume} {135}},\ \bibinfo {pages} {395} (\bibinfo {year} {2015})}\BibitemShut {NoStop}%
\bibitem [{\citenamefont {Blondel}\ \emph {et~al.}(2008)\citenamefont {Blondel}, \citenamefont {Guillaume}, \citenamefont {Lambiotte},\ and\ \citenamefont {Lefebvre}}]{blondel2008fast}%
  \BibitemOpen
  \bibfield  {author} {\bibinfo {author} {\bibfnamefont {V.~D.}\ \bibnamefont {Blondel}}, \bibinfo {author} {\bibfnamefont {J.-L.}\ \bibnamefont {Guillaume}}, \bibinfo {author} {\bibfnamefont {R.}~\bibnamefont {Lambiotte}},\ and\ \bibinfo {author} {\bibfnamefont {E.}~\bibnamefont {Lefebvre}},\ }\href {https://doi.org/10.1088/1742-5468/2008/10/P10008} {\bibfield  {journal} {\bibinfo  {journal} {Journal of Statistical Mechanics: Theory and Experiment}\ }\textbf {\bibinfo {volume} {2008}},\ \bibinfo {pages} {P10008} (\bibinfo {year} {2008})}\BibitemShut {NoStop}%
\bibitem [{\citenamefont {Simpson}\ and\ \citenamefont {Weiner}(1989)}]{oed}%
  \BibitemOpen
  \bibinfo {editor} {\bibfnamefont {J.~A.}\ \bibnamefont {Simpson}}\ and\ \bibinfo {editor} {\bibfnamefont {E.~S.~C.}\ \bibnamefont {Weiner}},\ eds.,\ \href@noop {} {\emph {\bibinfo {title} {The Oxford English Dictionary}}},\ \bibinfo {edition} {2nd}\ ed.\ (\bibinfo  {publisher} {Clarendon Press},\ \bibinfo {year} {1989})\BibitemShut {NoStop}%
\bibitem [{\citenamefont {Granger}(1969)}]{granger1969investigating}%
  \BibitemOpen
  \bibfield  {author} {\bibinfo {author} {\bibfnamefont {C.~W.~J.}\ \bibnamefont {Granger}},\ }\href@noop {} {\bibfield  {journal} {\bibinfo  {journal} {Econometrica}\ }\textbf {\bibinfo {volume} {37}},\ \bibinfo {pages} {424} (\bibinfo {year} {1969})}\BibitemShut {NoStop}%
\bibitem [{\citenamefont {Seabold}\ and\ \citenamefont {Perktold}(2010)}]{seabold2010statsmodels}%
  \BibitemOpen
  \bibfield  {author} {\bibinfo {author} {\bibfnamefont {S.}~\bibnamefont {Seabold}}\ and\ \bibinfo {author} {\bibfnamefont {J.}~\bibnamefont {Perktold}},\ }\href@noop {} {\bibfield  {journal} {\bibinfo  {journal} {Proceedings of the 9th Python in Science Conference}\ }\textbf {\bibinfo {volume} {57}},\ \bibinfo {pages} {61} (\bibinfo {year} {2010})}\BibitemShut {NoStop}%
\bibitem [{\citenamefont {Ladas}\ and\ \citenamefont {Qian}(1994)}]{Ladas1994}%
  \BibitemOpen
  \bibfield  {author} {\bibinfo {author} {\bibfnamefont {G.}~\bibnamefont {Ladas}}\ and\ \bibinfo {author} {\bibfnamefont {C.}~\bibnamefont {Qian}},\ }\href {https://doi.org/10.1080/02681119408806174} {\bibfield  {journal} {\bibinfo  {journal} {Dynamics and Stability of Systems}\ }\textbf {\bibinfo {volume} {9}},\ \bibinfo {pages} {153} (\bibinfo {year} {1994})}\BibitemShut {NoStop}%
\bibitem [{\citenamefont {{Google Books}}()}]{googlebooks}%
  \BibitemOpen
  \bibfield  {author} {\bibinfo {author} {\bibnamefont {{Google Books}}},\ }\href@noop {} {\bibinfo {title} {Google books ngram viewer}},\ \bibinfo {howpublished} {\url{https://books.google.com/ngrams}},\ \bibinfo {note} {accessed: 2022-03-02}\BibitemShut {NoStop}%
\bibitem [{\citenamefont {Davies}(2010{\natexlab{b}})}]{davies2010coha}%
  \BibitemOpen
  \bibfield  {author} {\bibinfo {author} {\bibfnamefont {M.}~\bibnamefont {Davies}},\ }\href@noop {} {\bibinfo {title} {The corpus of historical american english (coha): 400 million words, 1810–2009}},\ \bibinfo {howpublished} {\url{https://www.english-corpora.org/coha/}} (\bibinfo {year} {2010}{\natexlab{b}}),\ \bibinfo {note} {accessed: 2023-05-12}\BibitemShut {NoStop}%
\bibitem [{rep()}]{repo}%
  \BibitemOpen
  \href@noop {} {}\bibinfo {howpublished} {\url{https://github.com/AlePardoPintos/Words-cycles-economic.git}},\ \bibinfo {note} {accessed: 2025-07-10}\BibitemShut {NoStop}%
\end{thebibliography}%
\clearpage
\onecolumngrid
\section*{\large Supplementary Material}
\renewcommand{\thefigure}{S\arabic{figure}}
\renewcommand{\thetable}{S\arabic{table}}
\setcounter{figure}{0}
\setcounter{table}{0}
\newcounter{SIfig}
\renewcommand{\theSIfig}{S\arabic{SIfig}}

\begin{figure*}[ht]
\centering
\includegraphics[width=12cm]{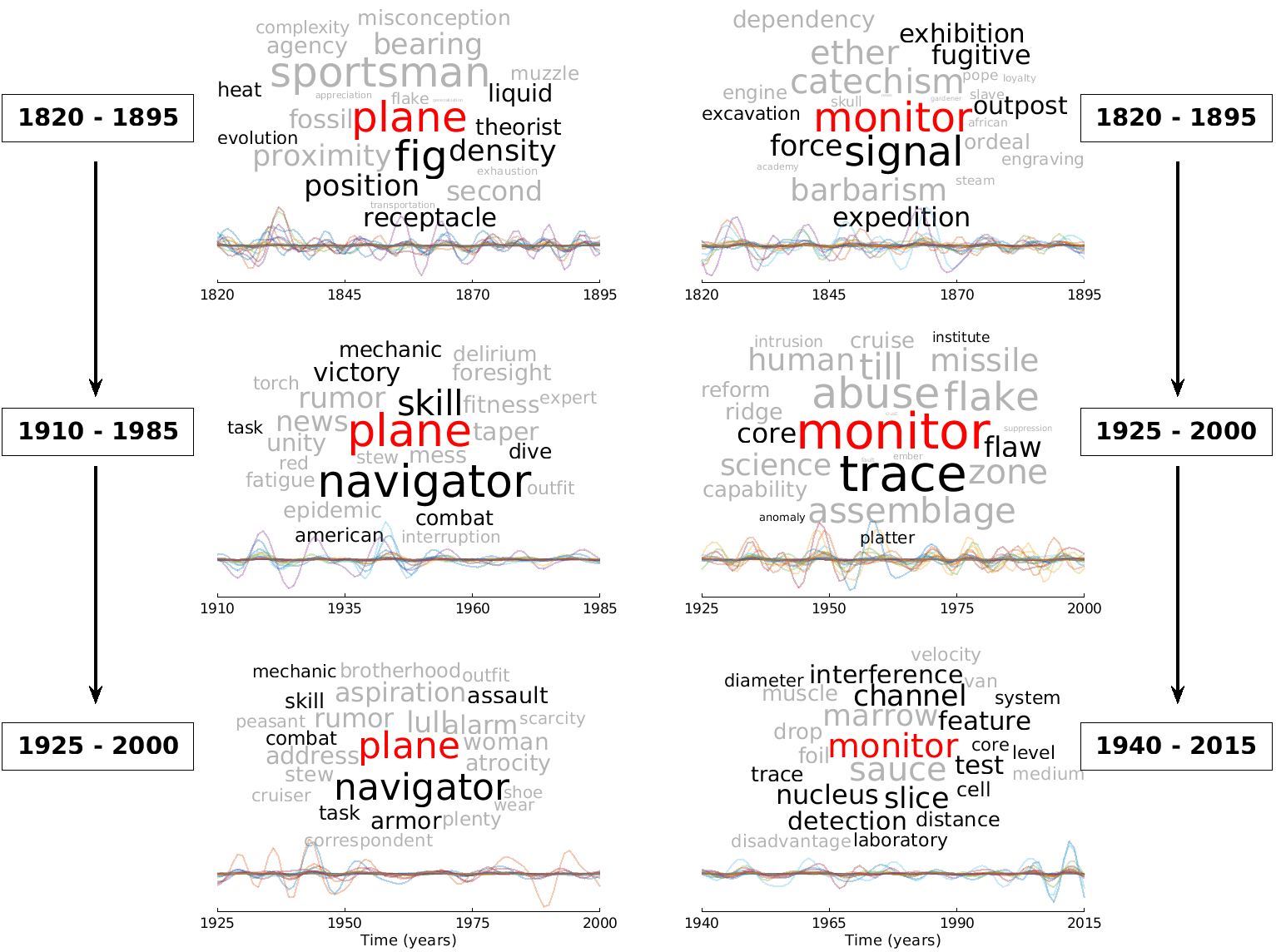}    
\caption{\textbf{Evolution of word communities and semantic shifts}. Each row shows the community containing the target word at a different times. Words in black refer to the dominant meaning of the target noun at that time. For the target \textit{plane}, the earliest community includes terms such as \textit{fig}, \textit{density}, \textit{liquid}, and \textit{position}, reflecting a mathematical context. Later communities shift toward an aeronautical meaning, with words like \textit{navigator}, \textit{skill}, \textit{mechanic}, \textit{american}, \textit{armor}, \textit{assault}, and combat. For \textit{monitor}, the earliest community includes \textit{signal}, \textit{force}, \textit{outpost}, and \textit{expedition}, consistent with its use as someone who issues alerts. Later communities include \textit{trace}, \textit{core}, \textit{channel}, \textit{system}, \textit{level}, and \textit{laboratory}, reflecting its modern use in the context of measuring or tracking outputs. These examples are included in Table \ref{table:meanings}, which were used to tune the Louvain resolution parameter, selecting values that ensure that words undergoing semantic shifts (like plane and monitor) are grouped with neighbors that reflect their meaning at each point in time.}
\refstepcounter{SIfig}\label{fig:communities_time}
\end{figure*}

\begin{figure*}[ht]
\centering
\includegraphics[width=15cm]{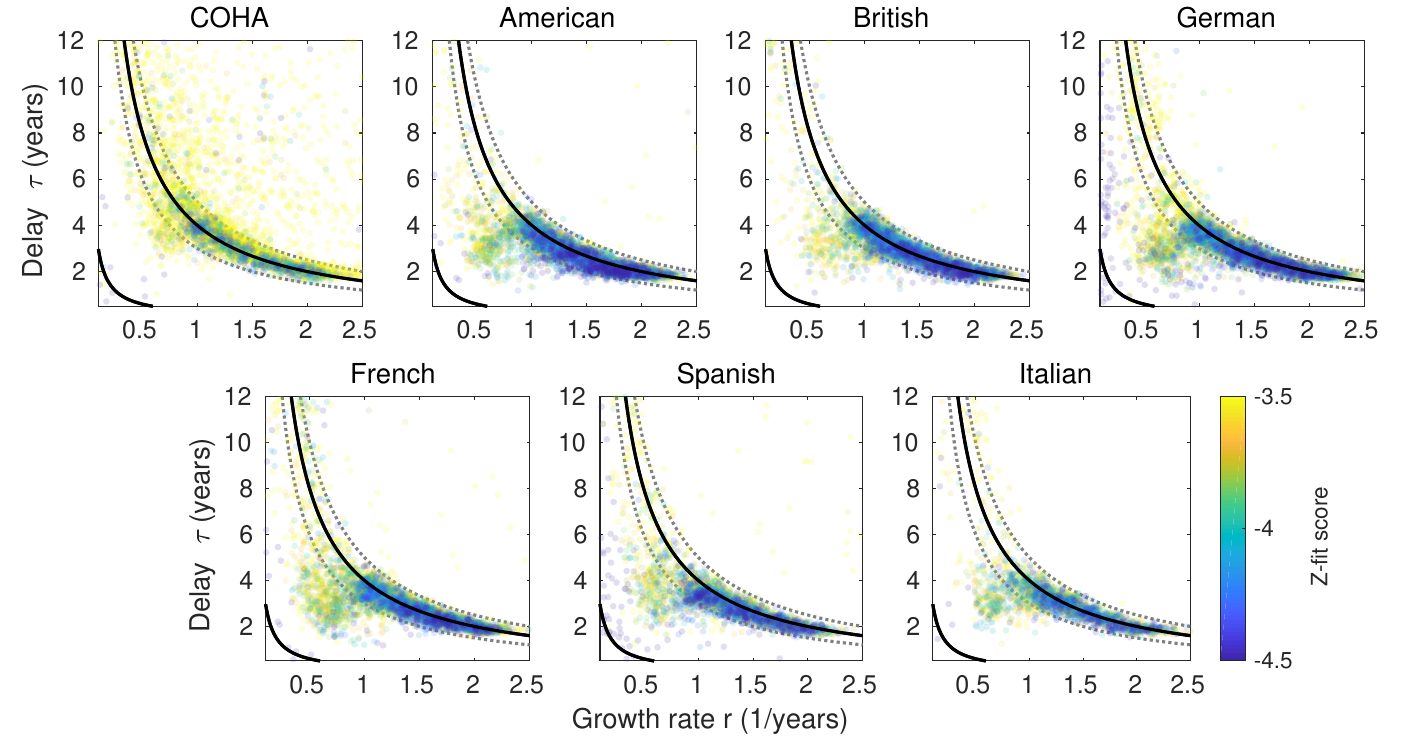}
\caption{\textbf{Parameter fitting is tuned across languages}. The logistic model with delay shows a Hopf bifurcation at $\tau = 4/r$ (black line). Individual fits for each language are shown, positioned around the bifurcation. The dashed lines indicate the Hopf neighborhood defined in Methods. Points are colored according to their Z-fit score, as defined in Methods. Lower or more negative values indicate better fits and are shown in blue. Mean fit parameters for each dataset are: COHA, $r = 1.3 \pm 0.6~1/\text{years}, \tau = 4 \pm 2$ years; American, $r = 1.4 \pm 0.5~1/\text{years}, \tau = 3 \pm 1$ years; British, $r = 1.4 \pm 0.4~1/\text{years}, \tau = 3 \pm 1$ years; German, $r = 1.3 \pm 0.5~1/\text{years}, \tau = 3 \pm 2$ years; French, $r = 1.3 \pm 0.5~1/\text{years}, \tau = 3 \pm 2$ years.}
\refstepcounter{SIfig}\label{fig:sup_hopf}
\end{figure*}

\begin{figure*}[ht]
\centering
\includegraphics[width=11cm]{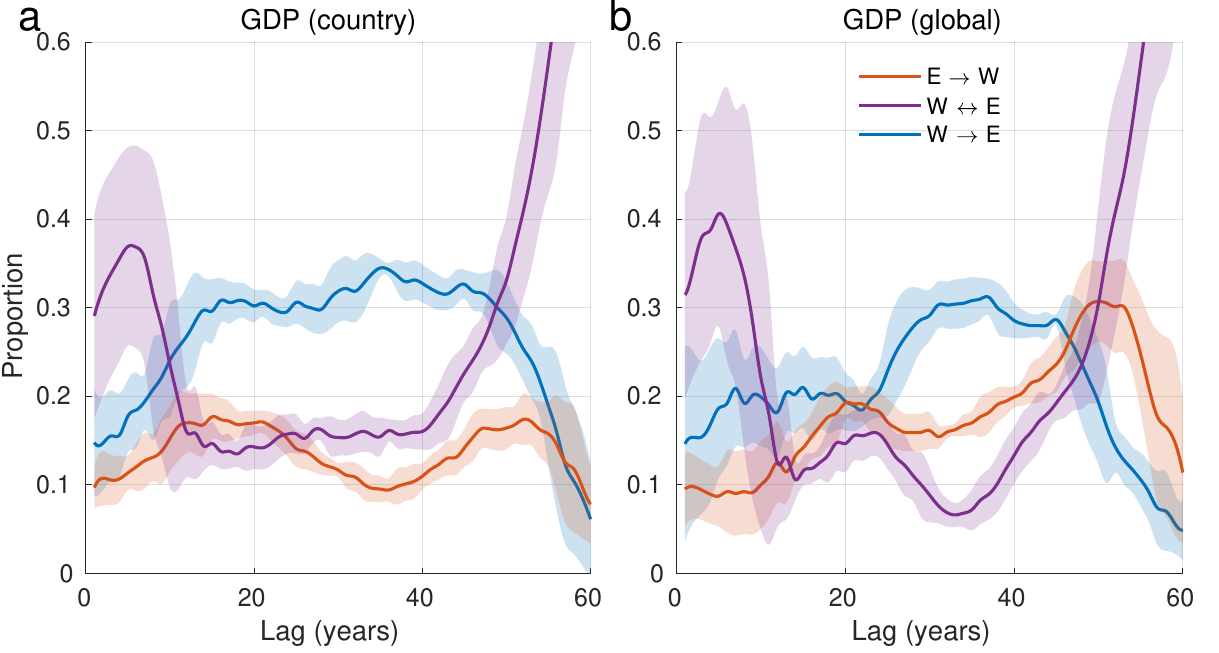}    
\caption{\textbf{Community proportions for each result of the Granger test as a function of the lag $k$}. When comparing the language series with GDP by country (a) or global GDP (b), it is observed that for lags between 20 and 40 years, the proportions remain constant, with GDP-driving being the most prominent direction. As the lag increases, the bilateral direction becomes more dominant. This is logical, as a lag $k$ includes all previous values, thereby encompassing the entire series.}
\refstepcounter{SIfig}\label{fig:lag_granger}
\end{figure*}

\begin{table}[ht]
\scriptsize
\begin{tabular}{|c|c|c|c|c|c|}
\hline
Word                     & Year                                                   & \begin{tabular}[c]{@{}c@{}}Earlier\\ meaning\end{tabular}                                                                                                                                             & Year & \begin{tabular}[c]{@{}c@{}}Present\\ meaning\end{tabular}                                                                                                                                             & Base                                                    \\ \hline
plastic                  & 1803                                                   & \begin{tabular}[c]{@{}c@{}}A substance that is easily moulded or \\ shaped under some conditions\\  but that solidifies as it cools, dries, etc.\end{tabular}                                         & 1909 & \begin{tabular}[c]{@{}c@{}}Any of a large and varied class of materials used\\  widely in manufacturing, which are organic\\  polymers of high molecular weight,\\  now usually based…\end{tabular} & \begin{tabular}[c]{@{}c@{}}COHA\\ American\end{tabular} \\ \hline
plane                    & 1604                                                   & \begin{tabular}[c]{@{}c@{}}A flat geometrical surface which has the property\\  that every straight line joining\\  any two points of the surface lies wholly in the surface.\end{tabular}            & 1908 & An aeroplane                                                                                                                                                                                        & \begin{tabular}[c]{@{}c@{}}COHA\\ American\end{tabular} \\ \hline
episode                  & 1678                                                   & \begin{tabular}[c]{@{}c@{}}In the Old Greek Tragedy, the interlocutory parts\\  between two choric songs,\\  because these were originally interpolations.\end{tabular}                               & 1915 & \begin{tabular}[c]{@{}c@{}}Each of the instalments into which a film, \\ television or radio drama, etc.\\  is divided for transmitting as a series.\end{tabular}                                   & \begin{tabular}[c]{@{}c@{}}COHA\\ American\end{tabular} \\ \hline
honey                    & \begin{tabular}[c]{@{}c@{}}Old \\ English\end{tabular} & \begin{tabular}[c]{@{}c@{}}A sweet sticky fluid or semi-solid substance from\\  whitish to dark brown in colour, produced by\\  honeybees, other social bees, and certain other…\end{tabular}         & 1843 & \begin{tabular}[c]{@{}c@{}}Now chiefly colloquial. Chiefly North American. \\ A good-looking or sexually attractive woman.\end{tabular}                                                             & COHA                                                    \\ \hline
\multirow{2}{*}{record}  & 1399                                                   & \begin{tabular}[c]{@{}c@{}}Anything preserving information and constituting\\  a piece of evidence about past events;\\  esp. an account kept in writing or some other permanent…\end{tabular}        & 1860 & \begin{tabular}[c]{@{}c@{}}The best performance or most remarkable \\ event of its kind; spec. the best officially \\ recorded achievemen of a particular \\ kind in a competitive…\end{tabular}    & \multirow{2}{*}{COHA}                                   \\ \cline{2-5}
                         & 1325                                                   & \begin{tabular}[c]{@{}c@{}}Law. The fact or condition of being, or of having been, \\ written down as evidence of a legal matter, esp.\\  as part of the proceedings or verdict of…\end{tabular}      & 1878 & \begin{tabular}[c]{@{}c@{}}Originally: a cylinder carrying a recording made\\  by a phonograph (now historical). \\ Later: a thin disc, latterly of plastic, \\ carrying recorded…\end{tabular}     &                                                         \\ \hline
album                    & 1612                                                   & \begin{tabular}[c]{@{}c@{}}A book in which contributions (such as signatures, \\ memorial verses, and epigrams)\\  are inscribed for the owner, esp. as \\ mementos or keepsakes. Also…\end{tabular}  & 1945 & \begin{tabular}[c]{@{}c@{}}A collection of recordings issued as a single\\  item on record, cassette, CD, etc.\end{tabular}                                                                         & COHA                                                    \\ \hline
channel                  & 1387                                                   & \begin{tabular}[c]{@{}c@{}}The hollow bed of a river, stream, \\ or other body of running water;\\  the course through which a river or stream flow\end{tabular}                                      & 1923 & \begin{tabular}[c]{@{}c@{}}A band of frequencies in the electromagnetic \\ spectrum, used for the transmission of a \\ radio or television signal.\end{tabular}                                     & COHA                                                    \\ \hline
joint                    & 1477                                                   & \begin{tabular}[c]{@{}c@{}}That wherein or whereby two component \\ members or elements of an artificial structure or \\ mechanism are joined \\ or fitted together, either so as to be…\end{tabular} & 1953 & Prison. U.S. slang.                                                                                                                                                                                 & COHA                                                    \\ \hline
\multirow{2}{*}{signal}  & 1413                                                   & \begin{tabular}[c]{@{}c@{}}An indication or token of a fact, quality, \\ future occurrence, etc.; a sign, a symbol. \\ Also as a mass noun. Frequently with of, that.\end{tabular}                    & 1838 & \begin{tabular}[c]{@{}c@{}}An alteration of an electric current, \\ electromagnetic wave, or the like by means of \\ which information is conveyed\\  from one place to another\end{tabular}        & \multirow{2}{*}{American}                               \\ \cline{2-5}
                         & 1576                                                   & \begin{tabular}[c]{@{}c@{}}A (usually prearranged) gesture, action,\\  or sound acting as the prompt for a particular action,\\  esp. a military manoeuvre.\end{tabular}                              & 1912 & \begin{tabular}[c]{@{}c@{}}A message communicated by radio;\\  a group of radio wavestransmitted or emitted \\ by a source and potentially detectable\end{tabular}                                  &                                                         \\ \hline
\multirow{2}{*}{monitor} & 1530                                                   & \begin{tabular}[c]{@{}c@{}}A school pupil or (esp. U.S.) college student assigned \\ disciplinary or other responsibilities\\  (formerly in some cases including teaching of junior\end{tabular}      & 1928 & \begin{tabular}[c]{@{}c@{}}A person who uses monitoring equipment\\  to check levels, standards, etc.\end{tabular}                                                                                  & \multirow{2}{*}{American}                               \\ \cline{2-5}
                         & 1623                                                   & A reminder or warning; a signal or indicator. Now rare.                                                                                                                                               & 1931 & \begin{tabular}[c]{@{}c@{}}Something which monitors or displays \\ performance, output, etc., esp. of a system.\\  A cathode ray tube screen\\  or other device used to monitor…\end{tabular}       &                                                         \\ \hline
\end{tabular}
\caption{Nouns used to validate the linguistic relevance of the computed community structure. For each noun, we list the original meaning and its year of entry in the Oxford English Dictionary, along with the new meaning and the year it was added.}
\label{table:meanings}
\end{table}

\clearpage

\begin{longtable}[ht]{|p{1.1cm}|c|p{13.4cm}|p{3cm}|}
\hline
\textbf{Lang} & \textbf{N} & \textbf{Main Words} & \textbf{Semantic field} \\
\hline
\endfirsthead
\hline
\textbf{Lang} & \textbf{N} & \textbf{Main Words} & \textbf{Semantic field} \\
\hline
\endhead

\multirow{18}{*}{USA} & 1 & state, case, law, section, company, court, act, person, information, effect & Law \\\cline{2-4}
 & 2 & system, number, use, study, group, value, result, form, data, process & Science \\
\cline{2-4}
 & 3 & eye, friend, mother, face, moment, love, voice, morning, rest, fear & Emotion \\
\cline{2-4}
 & 4 & heart, passion, pride, fate, affection, midst, triumph, throne, multitude, kindness & Emotion \\
\cline{2-4}
 & 5 & machine, box, minute, steel, plate, speed, block, edge, pipe, cut & Science \\
\cline{2-4}
 & 6 & force, enemy, soldier, camp, battle, captain, campaign, guard, gun, troop & Military \\
\cline{2-4}
 & 7 & son, king, lord, truth, heaven, virtue, temple, greek, roman, priest & Religion \\
\cline{2-4}
 & 8 & service, order, government, committee, department, president, war, officer, effort, supply & Government \\
\cline{2-4}
 & 9 & condition, patient, treatment, disease, blood, medicine, skin, tissue, bone, muscle & Medicine \\
\cline{2-4}
 & 10 & line, point, side, foot, wall, length, inch, spring, distance, iron & Science \\
\cline{2-4}
 & 11 & product, price, plan, industry, production, market, trade, need, oil, gas & Economy \\
\cline{2-4}
 & 12 & money, cause, good, circumstance, debt, entry, possession, settlement, favor, deed & Law \textendash Government \\
\cline{2-4}
 & 13 & god, sin, happiness, flesh, prophet, devil, scripture, ghost, salvation, disciple & Religion \\
\cline{2-4}
 & 14 & parliament, ambition, zeal, siege, fury, resentment, recourse, tyranny, sincerity, calamity & Emotion\textendash Social \\
\cline{2-4}
 & 15 & position, division, army, communication, staff, front, command, leave, corps, transport & Military \\
\cline{2-4}
 & 16 & plant, center, agency, equipment, job, worker, facility, training, transportation, engineer & Economy \textendash Government \\
\cline{2-4}
 & 17 & soil, species, culture, variety, fruit, seed, root, leaf, grass, insect & Domestic \textendash Nature \\
\cline{2-4}
 & 18 & people, nation, peace, help, necessity, destruction, coffee, uniform, victor, hostility & Military \\
\cline{1-4}
\multirow{22}{*}{British} & 1 & case, state, country, number, government, power, point, question, service, member & Economy \textendash Government \\\cline{2-4}
 & 2 & hand, eye, room, night, face, moment, door, voice, morning, girl & Body \textendash Domestic \\
\cline{2-4}
 & 3 & effect, group, study, change, process, theory, difference, surface, science, structure & Science \\
\cline{2-4}
 & 4 & war, force, officer, general, army, treatment, supply, east, peace, president & Military \\
\cline{2-4}
 & 5 & hope, feel, conversation, passion, imagination, coffee, uncle, beast, fool, curiosity & Emotion \textendash Social \\
\cline{2-4}
 & 6 & mind, father, heart, mother, lady, love, sister, thought, joy, silence & Emotion \\
\cline{2-4}
 & 7 & use, result, problem, cent, method, material, air, practice, test, instance & Economy \\
\cline{2-4}
 & 8 & god, spirit, christian, faith, heaven, understanding, doctrine, grace, sin, prayer & Religion \\
\cline{2-4}
 & 9 & company, account, section, office, council, evidence, march, date, public, return & Law \\
\cline{2-4}
 & 10 & form, example, nature, idea, sense, knowledge, existence, reality, interpretation, distinction & Philosophy \\
\cline{2-4}
 & 11 & king, death, wife, brother, horse, queen, daughter, marriage, duke, husband & Military \textendash Religion \\
\cline{2-4}
 & 12 & temperature, metal, tube, edge, sheet, pipe, bottom, hole, shaft, cylinder & Science \\
\cline{2-4}
 & 13 & house, lord, right, bill, measure, gentleman, parliament, argument, speech, amendment & Government \\
\cline{2-4}
 & 14 & term, rule, building, district, provision, railway, share, agreement, construction, site & Economy \textendash Law \\
\cline{2-4}
 & 15 & hour, patient, disease, blood, skin, brain, tissue, muscle, symptom, nerve & Medicine \\
\cline{2-4}
 & 16 & act, law, person, court, justice, judgment, trial, judge, proceeding, offense & Law \\
\cline{2-4}
 & 17 & direction, frame, wheel, apparatus, cover, specification, thread, screw, lever, nut & Science \\
\cline{2-4}
 & 18 & music, song, opera, concert, harmony, composer, singer, piano, musician, orchestra & Culture \\
\cline{2-4}
 & 19 & evening, hat, guest, match, mistress, breakfast, glance, curtain, penny, laughter & Body \textendash Domestic \\
\cline{2-4}
 & 20 & system, development, research, management, region, aspect, organization, agency, identity, category & Philosophy \textendash Law \\
\cline{2-4}
 & 21 & character, english, writer, style, writing, version, essay, verse, reflection, sentiment & Culture \\
\cline{2-4}
 & 22 & way, friend, home, doubt, sort, chance, lot, bit, spite, mistake & Emotion \textendash Military \\
\cline{1-4}
\multirow{10}{*}{COHA} & 1 & day, way, hand, thing, eye, woman, house, face, head, place & Body \textendash Domestic \\\cline{2-4}
 & 2 & part, state, fact, question, law, point, interest, nature, system, view & Philosophy \textendash Law \\
\cline{2-4}
 & 3 & country, power, government, right, nation, period, opinion, policy, authority, congress & Government \\
\cline{2-4}
 & 4 & egg, milk, meat, sugar, salt, chicken, dish, vegetable, cake, butter & Domestic \\
\cline{2-4}
 & 5 & hour, wind, master, cloud, leaf, darkness, breast, gate, brow, kiss & Nature \\
\cline{2-4}
 & 6 & price, material, method, industry, cost, production, product, supply, resource, proportion & Economy \\
\cline{2-4}
 & 7 & war, force, enemy, battle, attack, troop, command, victory, advance, destruction & Military \\
\cline{2-4}
 & 8 & friend, mind, think, sight, fear, feel, pleasure, thought, secret, affection & Emotion \\
\cline{2-4}
 & 9 & study, student, college, university, education, degree, teacher, skill, training, instruction & Social \\
\cline{2-4}
 & 10 & stone, edge, tooth, coffee, pair, tea, tip, diamond, glove, veil & Objects \\
\cline{1-4}
\pagebreak
\multirow{28}{*}{German} & 1 & mensch, gott, leben, welt, wort, natur, denken, sachen, worten, dingen & Emotion \textendash Philosophy \\\cline{2-4}
 & 2 & mannen, hand, hausen, augen, weg, kopfen, blicken, muttern, vatern, sichten & Body \textendash Domestic \\
\cline{2-4}
 & 3 & fallen, fällen, Anwendung, Behandlung, krankheit, grad, verlaufen, wechen, patient, vorgehen & Medicine \\
\cline{2-4}
 & 4 & beispiel, system, unterscheiden, darstellung, veränderung, unternehmen, element, funktion, eigenschaft, beziehen & Economy \\
\cline{2-4}
 & 5 & gründen, Entscheidung, annahme, kosten, verfügung, ausnehmen, Anspruch, schutzen, bestimmung, antragen & Law \\
\cline{2-4}
 & 6 & wassern, wirkung, luften, drucken, temperatur, Minute, farbe, eisen, flüssigkeit, gasen & Science \\
\cline{2-4}
 & 7 & bedeutung, versuchen, beziehung, zahlen, reihen, untersuchengen, untersuchung, vergleichen, gruppe, anzahl & Science \\
\cline{2-4}
 & 8 & frau, stimmen, mädchen, tür, sonnen, feuern, hund, weinen, bäumen, weisen & Emotion \textendash Nature \\
\cline{2-4}
 & 9 & recht, staat, gesetz, partei, politik, mitglied, Staat, meinung, forderung, zukunft & Government \\
\cline{2-4}
 & 10 & werken, kirchen, Schrift, christus, christi, Kirche, priest, jesu, paulus, glaubens & Religion \\
\cline{2-4}
 & 11 & teil, wert, hilfe, einflußen, werten, maßnehmen, änderung, mitteilung, kontroll, Beurteilung & Economy \textendash Government \\
\cline{2-4}
 & 12 & theil, theile, werth, größ, mittheilung, verhältniss, thieren, thatsache, vortheil, thatsach & Economy \\
\cline{2-4}
 & 13 & sinnen, kritik, Erfahrung, wirklichkeit, Vorstellung, existenz, sätzen, bewußtsein, kanten, individuum & Philosophy \\
\cline{2-4}
 & 14 & richtung, längen, flächen, gewichten, Linie, wellen, winkeln, durchmessern, kugeln, intensität & Science \\
\cline{2-4}
 & 15 & ausdrucken, wissenschaft, charakter, moment, ganz, philosophie, religion, idee, erkenntnis, persönlichkeit & Philosophy \\
\cline{2-4}
 & 16 & machen, ordningen, freiheit, willen, Gewalt, opfern, geschäft, widerspruch, einzeln, schwächen & Government \textendash Social \\
\cline{2-4}
 & 17 & begreifen, gegensatz, einheiten, wissen, beweisen, ursprung, Anschauung, identität, materie, wissens & Science \textendash Philosophy \\
\cline{2-4}
 & 18 & preis, schwierigkeit, monat, verwendung, umfangen, heften, betrachten, Einrichtung, Erhöhung, ix & Economy \textendash Law \\
\cline{2-4}
 & 19 & betreiben, anteil, gründung, Berbesserung, west, nutzung, japan, versorgung, beseitigung, schaffungen & Economy \textendash Government \\
\cline{2-4}
 & 20 & musik, werk, publikum, theater, bühne, oper, chor, komponist, Aufführung, melodie & Culture \\
\cline{2-4}
 & 21 & kunst, künstl, füllen, poesie, schaffen, vollendung, malerei, kun, kunstwerk, zeitgenossen & Culture \\
\cline{2-4}
 & 22 & mai, august, juni, märz, juli, april, september, november, januar, oktober & Social \\
\cline{2-4}
 & 23 & mal, eltern, schmerzen, essen, nase, hitzen, abends, katz, giften, julia & Domestic \\
\cline{2-4}
 & 24 & ddr, konflikt, gmbh, aspekt, struktur, kontext, perspektive, umwelt, cdu, entwicklingen & Government \textendash Social \\
\cline{2-4}
 & 25 & feststellung, Hinweis, veröffentlichung, ergänzungen, einnehmen, Geitraum, auskunft, unterliegen, wortlauten, Schaden & Government \textendash Culture \\
\cline{2-4}
 & 26 & steinen, sanden, fluß, gebirge, gestein, formation, fluss, jura, harzen, kreiden & Science \\
\cline{2-4}
 & 27 & substanz, nerv, entzündung, bluten, stadien, drüsen, Membran, kesseln, iris, embryo & Medicine \textendash Science \\
\cline{2-4}
 & 28 & feind, hafen, posen, offizieren, kilomet, mannschaft, beförderung, norwegen, wehren, meldungen & Military \\
 \cline{1-4}
 \multirow{30}{*}{French} & 1 & homme, main, œil, tête, peine, son, coup, cœur, pied, toi & Body \textendash Emotion \\\cline{2-4}
 & 2 & france, guerre, force, armée, nation, paix, feu, effort, empire, allemagne & Military \\
\cline{2-4}
 & 3 & dieu, esprit, fils, grâce, parole, frère, vérité, seigneur, foi, christ & Religion \\
\cline{2-4}
 & 4 & vie, fois, chose, idée, amour, pensée, sentiment, mal, souvenir, passion & Culture \textendash Emotion \\
\cline{2-4}
 & 5 & droit, loi, article, service, conseil, intérêt, mois, ministre, titre, membre & Law \\
\cline{2-4}
 & 6 & travail, effet, rapport, condition, cours, système, résultat, problème, développement, base & Economy \\
\cline{2-4}
 & 7 & regard, soir, impression, bruit, fenêtre, garçon, coin, sourire, robe, cuisine & Domestic \\
\cline{2-4}
 & 8 & élément, type, niveau, volume, formation, surface, distance, pression, procédé, gaz & Science \\
\cline{2-4}
 & 9 & année, faire, exemple, science, origine, rôle, construction, mise, association, critique & Science \textendash Government \\
\cline{2-4}
 & 10 & ordre, principe, peuple, pouvoir, liberté, justice, opinion, révolution, instruction, constitution & Government \\
\cline{2-4}
 & 11 & forme, fig, feuille, longueur, contact, dessous, sommet, tissu, tube, fil & Science \\
\cline{2-4}
 & 12 & pays, question, gouvernement, mesure, compte, projet, situation, commission, politique, population & Government \\
\cline{2-4}
 & 13 & jour, heure, moment, chef, sang, devoir, matin, semaine, journée, lendemain & Emotion \textendash Domestic \\
\cline{2-4}
 & 14 & honneur, volonté, crime, respect, victime, violence, ruine, espérance, haine, promesse & Emotion \\
\cline{2-4}
 & 15 & tribunal, jugement, juge, arrêt, motif, obligation, propriétaire, exception, immeuble, dommage & Law \\
\cline{2-4}
 & 16 & eau, quantité, poids, température, transformation, variation, mélange, huile, détermination, intensité & Science \\
\cline{2-4}
 & 17 & père, mourir, maître, religion, don, sœur, baron, dessein, pauvre, amitié & Religion \textendash Social \\
\cline{2-4}
 & 18 & roi, prince, duc, comte, charles, cardinal, chevalier, couronne, orléans, monseigneur & History \\
\cline{2-4}
 & 19 & structure, référence, ailleurs, comportement, coût, contexte, critère, orientation, norme, salarié & Economy \\
\cline{2-4}
 & 20 & raison, sens, nature, connaissance, existence, expression, réalité, philosophie, doctrine, langage & Philosophy \\
\cline{2-4}
 & 21 & côté, face, bord, dessus, peau, bande, arrière, dent, pointe, trou & Body \textendash Domestic \\
\cline{2-4}
 & 22 & sieur, religieux, bourgogne, diocèse, bourg, abbaye, chanoine, denier, clerc, dijon & Religion \\
\cline{2-4}
 & 23 & vue, tableau, modèle, changement, appareil, courir, vapeur, précision, moteur, dispositif & Science \\
\cline{2-4}
 & 24 & œuvre, trait, goût, public, lecteur, littérature, écrivain, mérite, talent, spectacle & Culture \\
\cline{2-4}
 & 25 & acide, excès, cuivre, sel, plomb, chaux, sels, liqueur, potasse, chlorure & Science \\
\cline{2-4}
 & 26 & histoire, livre, auteur, ouvrage, texte, mémoire, vol, édition, in, eglise & History \textendash Culture \\
\cline{2-4}
 & 27 & saint, siècle, jean, pierre, marier, françois, jacque, henri, robert, recueil & History \textendash Social \\
\cline{2-4}
 & 28 & milieu, air, fond, morceau, course, bain, vide, glace, flamme, appartement & Domestic \\
\cline{2-4}
 & 29 & état, égard, port, navire, marchandise, étranger, avance, taxe, destination, payement & Economy \\
\cline{2-4}
 & 30 & ville, arme, capitaine, cents, arrivée, garnison, fuite, permission, guerrier, cavalier & Military \\
\cline{1-4}
\multirow{20}{*}{Spanish} & 1 & hombre, vida, casa, mundo, mujer, mano, padre, verdad, noche, paso & Domestic \textendash Emotion \\\cline{2-4}
 & 2 & tiempo, rey, muerte, amor, hermano, cielo, esperanza, suerte, tú, fuego & Religion \\
\cline{2-4}
 & 3 & ley, caso, derecho, efecto, acuerdo, término, recurso, acto, fecha, conocimiento & Law \\
\cline{2-4}
 & 4 & estado, país, relación, sistema, poder, desarrollo, interés, situación, actividad, necesidad & Government \\
\cline{2-4}
 & 5 & señor, gobierno, artículo, presidente, razón, proyecto, clase, comisión, ministro, fondo & Government \\
\cline{2-4}
 & 6 & momento, ojo, cabeza, sangre, mal, brazo, cara, error, boca, piel & Body \textendash Medicine \\
\cline{2-4}
 & 7 & josé, mes, francisco, maría, julio, antonio, mayo, diciembre, manuel, junio & Law \textendash Social \\
\cline{2-4}
 & 8 & agua, lado, mar, color, piedra, distancia, madera, cabo, extremo, segundo & Nature \textendash Body \\
\cline{2-4}
 & 9 & lugar, juan, pedro, doña, villa, vecino, obispo, diego, enrique, alfonso & Nature \textendash Social \\
\cline{2-4}
 & 10 & fuerza, jefe, general, ejército, arma, enemigo, soldado, tropa, marcha, golpe & Military \\
\cline{2-4}
 & 11 & superficie, contacto, frecuencia, presión, pared, aparato, tamaño, capa, temperatura, tejido & Medicine \textendash Science \\
\cline{2-4}
 & 12 & trabajo, organización, educación, programa, creación, participación, reunión, distribución, protección, alumno & Social \textendash Government \\
\cline{2-4}
 & 13 & contar, título, documento, código, siguiente, acta, supremo, cuenta, fomento, jurisdicción & Law \textendash Government \\
\cline{2-4}
 & 14 & real, plaza, letra, barcelona, torre, continuación, sevilla, valencia, ayuntamiento, zaragoza & Culture \textendash Social \\
\cline{2-4}
 & 15 & hijo, francia, reino, fernando, felipe, reina, castilla, conde, corona, duque & History \textendash Government \\
\cline{2-4}
 & 16 & arte, poeta, literatura, poesía, novela, personaje, crítica, escritor, lenguaje, lector & Culture \\
\cline{2-4}
 & 17 & españa, gente, ocasión, caballo, batalla, francés, italia, castillo, fortuna, fama & Domestic \textendash Government \\
\cline{2-4}
 & 18 & forma, producto, comercio, mercado, industria, grado, beneficio, tasa, costo, consumo & Economy \\
\cline{2-4}
 & 19 & hora, final, mañana, cuarto, minuto, habitación, conversación, pareja, tienda, hotel & Body \textendash Domestic \\
\cline{2-4}
 & 20 & peligro, temor, felicidad, ruina, promesa, talento, orgullo, ignorancia, terror, horror & Emotion \\
 \cline{1-4}
 \multirow{19}{*}{Italian} & 1 & parte, caso, diritto, articolo, atto, termine, causa, effetto, corte, norma & Law \\\cline{2-4}
 & 2 & tempo, volta, uomo, parola, mano, padre, occhio, voce, dire, cuore & Body \textendash Emotion \\
\cline{2-4}
 & 3 & essere, legge, giorno, ora, ordine, giorni, consiglio, servizio, mese, decreto & Economy \textendash Government \\
\cline{2-4}
 & 4 & cosa, vita, mondo, terra, morire, amore, amico, mente, notte, sentimento & Emotion \textendash Philosophy \\
\cline{2-4}
 & 5 & casa, donna, figlio, piede, ragazzo, giovane, moglie, faccia, sera, trarre & Domestic \\
\cline{2-4}
 & 6 & arte, poeta, poesia, scrittore, immagine, scena, personaggio, stile, verso, lettore & Culture \\
\cline{2-4}
 & 7 & cagione, cavallo, lorare, cavaliere, grandezza, alcuno, cavalli, tale, bontà, poiché & Military \textendash Religion \\
\cline{2-4}
 & 8 & ii, giovanni, napoli, francesco, carlo, maria, Pietro, antonio, del, bologna & Social \\
\cline{2-4}
 & 9 & forma, sistema, ricerca, tipo, lato, origine, metodo, sostanza, fenomeno, parete & Medicine \textendash Science \\
\cline{2-4}
 & 10 & città, re, francia, regno, romano, papa, principe, duca, compagnia, francese & History \textendash Government \\
\cline{2-4}
 & 11 & luce, aria, scala, collo, tono, solito, cane, pelle, filo, piccolo & Law \\
\cline{2-4}
 & 12 & idea, pensiero, realtà, spirito, verità, movimento, coscienza, linguaggio, tendenza, morale & Philosophy \\
\cline{2-4}
 & 13 & segno, malattia, paziente, alterazione, tumore, dose, clinica, lesioni, febbre, sintomo & Medicine \\
\cline{2-4}
 & 14 & opera, secolo, figura, palazzo, pietra, monumento, pittura, pittore, statua, cappella & Culture \\
\cline{2-4}
 & 15 & de, et, in, est, cum, ut, quod, loco, sunt, ni & History \\
\cline{2-4}
 & 16 & governo, ministro, camera, riforma, membro, voto, affare, discussione, rappresentante, comitato & Government \\
\cline{2-4}
 & 17 & guerra, pace, arma, nemico, fuoco, battaglia, pericolo, vittoria, esercito, comando & Military \\
\cline{2-4}
 & 18 & onda, maniera, colore, egli, costume, greco, antico, fama, ingegno, vestire & Domestic \\
\cline{2-4}
 & 19 & forza, bisogno, fronte, dovere, sforzo, vittima, pane, giornali, migliaio, miseria & Government \\
 \hline
\caption*{ }

\label{table:communities}
\end{longtable}
\vspace{-0.8cm}
\noindent TABLE S2: \textbf{Main words of the largest communities across databases.} Each community is annotated with its corresponding semantic domain.




\end{document}